\PassOptionsToPackage{table,xcdraw}{xcolor}
\documentclass[sigconf, manuscript]{acmart}

\usepackage{bm}
\usepackage{color}
\usepackage{tabularray}
\usepackage{longtable}
\usepackage{tabularx}
\usepackage{multirow}
\usepackage{subcaption}
\usepackage{ragged2e}
\usepackage{tabularx}
\usepackage{amsmath}

\usepackage{siunitx}
\renewcommand{\qty}[2]{\ensuremath{#1}}

\copyrightyear{2024}
\acmYear{2024}
\setcopyright{rightsretained}
\acmConference[CUI '24]{Conversational User Interfaces}{July 08--10, 2024}{Luxembourg City, Luxembourg}
\acmBooktitle{ (CUI '24), July 08--10, 2024, Luxembourg City, Luxembourg}
\acmDOI{XXXXXXX.XXXXXXX}
\acmISBN{978-1-4503-XXXX-X/18/06}

\begin{document}


\title{Examining Humanness as a Metaphor to Design Voice User Interfaces}


\author{Smit Desai}
\orcid{}
\affiliation{%
  \institution{University of Illinois at Urbana-Champaign}
  \country{USA}}
\email{smitad2@illinois.edu}

\author{Mateusz Dubiel}
\orcid{0000-0001-8250-3370}
\affiliation{%
  \institution{University of Luxembourg}
  \country{Luxembourg}}
\email{mateusz.dubiel@uni.lu}

\author{Luis A. Leiva}
\orcid{0000-0002-5011-1847}
\affiliation{%
  \institution{University of Luxembourg}
  \country{Luxembourg}}
\email{name.surname@uni.lu}


\begin{abstract}
Voice User Interfaces (VUIs) increasingly leverage `humanness' as a foundational design metaphor, adopting roles like `assistants,' `teachers,' and `secretaries' to foster natural interactions. Yet, this approach can sometimes misalign user trust and reinforce societal stereotypes, leading to socio-technical challenges that might impede long-term engagement. This paper explores an alternative approach to navigate these challenges—incorporating non-human metaphors in VUI design. We report on a study with 240 participants examining the effects of human versus non-human metaphors on user perceptions within health and finance domains. Results indicate a preference for the human metaphor (doctor) over the non-human (health encyclopedia) in health contexts for its perceived enjoyability and likeability. In finance, however, user perceptions do not significantly differ between human (financial advisor) and non-human (calculator) metaphors. Importantly, our research reveals that the explicit awareness of a metaphor's use influences adoption intentions, with a marked preference for non-human metaphors when their metaphorical nature is not disclosed. These findings highlight context-specific conversation design strategies required in integrating non-human metaphors into VUI design, suggesting tradeoffs and design considerations that could enhance user engagement and adoption.
\end{abstract}



\begin{teaserfigure}
\centering
  \includegraphics[width=\textwidth]{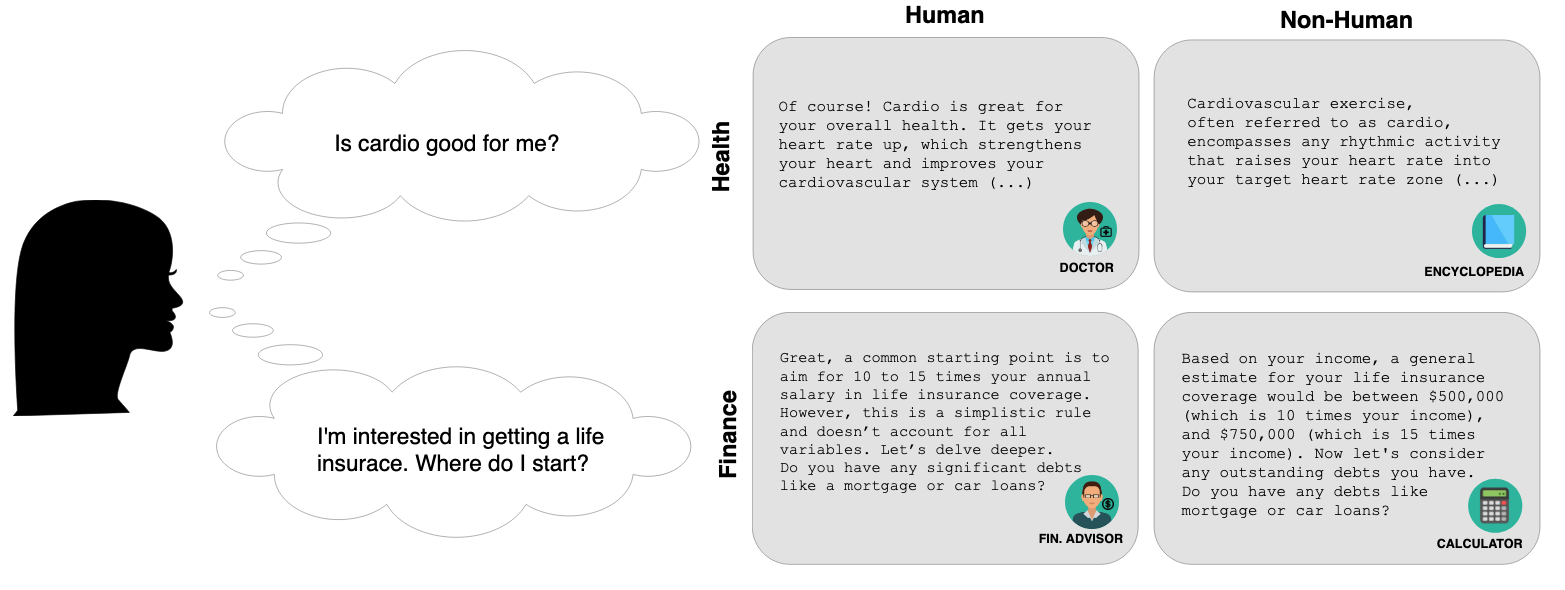}
  \caption{We explored how frequent users of Voice User Interfaces (VUIs) perceive interacting with human metaphors versus non-human metaphors in the domain of health and finance. For health, we designed two metaphorical VUIs that acted as a `doctor' or a `health encyclopedia.' For finance, we used the metaphors `financial advisor' and `calculator.'} 
  \label{fig:teaser}
\end{teaserfigure} 

\maketitle

\section{Introduction}\label{introduction}

Voice User Interfaces (VUIs), including voice assistants like Amazon Alexa or Apple Siri, have become ubiquitous in everyday life. They are used to perform various tasks like information seeking, home automation, playing games, or even for social companionship \cite{Sciuto_Saini_Forlizzi_Hong_2018}. Typically, they are designed to be and act as ``assistants'' \cite{Turk_2016}. However, in more specialized application domains like health and finance, they can be given the personalities of  ``teachers'' \cite{Desai_Chin_2023}, ``brokers'' \cite{Lee_Frank_IJsselsteijn_2021}, ``therapists'' \cite{Motalebi_Cho_Sundar_Abdullah_2019}, ``exercise coaches'' \cite{Desai_Hu_Lundy_Chin_2023}, or even ``financial advisors'' \cite{Zhu_Pysander_Soderberg_2023}.  These personalities, which act as the front-end conversational user interfaces (CUIs), are referred to as ``personas,'' and the act of designing personas acts as the starting point of the CUI design process \cite{Sadek_Calvo_Mougenot_2023}.   

CUIs, from ELIZA to ChatGPT, have been anthropomorphized due to their opaque black-box nature \cite{liao2023ai}. In most cases, the anthropomorphism in users originates from ``overlearned social behaviors'' \cite{Nass_Moon_2000}. However, in the case of VUIs, eliciting anthropomorphic tendencies is a conscious design decision \cite{Desai_Twidale_2023}. Conversation design guidelines of Google\footnote{https://developers.google.com/assistant/conversation-design/create-a-persona} and Amazon\footnote{https://developer.amazon.com/en-US/blogs/alexa/alexa-skills-kit/2018/08/hear-it-from-a-skill-builder-how-to-create-a-persona-for-your-alexa-skill} provide detailed instructions on designing personas, including picking emotional reactions and personality traits. Due to these practices, `humanness' has become the foundational metaphor of human-VUI interactions \cite{10.1145/3098279.3098539}. 

In recent literature, there has been much criticism of this foundational humanness metaphor for VUI design, highlighting issues from practical functionality to ethical considerations. From a practical perspective, presenting a VUI as an assistant results in higher pre-use expectations \cite{Khadpe_Krishna_Fei-Fei_Hancock_Bernstein_2020}, which, if unmet, results in infrequent use \cite{10.1145/3098279.3098539} or non-use \cite{10.1145/3322276.3322332}. 
Additionally, VUIs utilize a request-response mechanism that could be effective for executing command-oriented tasks but falls short of facilitating genuine `conversations' \cite{Porcheron_Fischer_Reeves_Sharples_2018} and thus fall significantly short of the human-human communication benchmark. Moreover, finding common ground with a machine (especially in the event of conversational errors) pretending to be a human can be especially frustrating \cite{Desai_Twidale_2022, Edlund_2019}. However, these frustrations pale compared to the ethical issues presented by the humanness metaphor. VUIs, designed to be assistants, are given a default female voice. Thereby, reinforcing existing societal stereotypes. In other cases, they are designed to act as "servants" or "secretaries," acting as ``invisible women'' in our houses \cite{Turk_2016}. These and other ethical and functional concerns have been well documented in previous CUI conferences \cite{Edlund_2019, McMillan_Jaber_2021, Pradhan_Lazar_2021, Simpson_Crone_2022, Desai_Twidale_2022}. 

Due to these concerns, there is a growing body of work proposing designing conversational interfaces using non-human metaphors. Until now, this research has been limited to chat-based CUIs, with the most notable example of \cite{Jung_Qiu_Bozzon_Gadiraju_2022}. Preliminary evidence (albeit for chatbots) suggests that non-human metaphors (e.g., animal, plant, object) could be useful in increasing user engagement and perceived workload \cite{Jung_Qiu_Bozzon_Gadiraju_2022} in comparison to human metaphors. Given the problematic nature of humanness metaphors in VUI design, it is important to explore how designing non-human metaphors could impact user perceptions. Especially considering there is evidence in the literature suggesting that users already do conceptualize VUIs as ``things'' \cite{10.1145/3322276.3322332} and routinely use non-human metaphors (like Genie, Encyclopedia, Tools) to describe them.  

To the best of our knowledge, there is no existing research that examines the comparison between VUIs designed with non-human metaphors and those using human metaphors in terms of user perceptions and willingness to adopt these technologies. To address this gap, we conduct a mixed-subjects study (N=240) to understand how frequent users of VUIs perceive interacting with a VUI named `Z,' designed using the metaphors `doctor' and `health encyclopedia' in the domain of health and 'financial advisor' and `calculator' in finance. Unlike chat-based conversational interfaces, which can signal their metaphor to users through avatars or icons, VUIs lack such clear affordances. Therefore, the approach involves either explicitly informing users about the underlying metaphor or leaving it implicit. To explore this, we also investigate how the metaphor's conveyance influences user perceptions, comparing the explicit versus implicit communication strategies. We operationalize the explicit versus implicit knowledge of the metaphor as awareness levels—with explicit awareness indicating that the participant is made aware of the metaphor used in the design process and implicit awareness signaling that the participants are not informed of the metaphor. With this paper, we explore two main research questions (RQs): 

\begin{itemize}
    \item \textbf{RQ1:} \textit{How do users of VUIs perceive interacting with non-human metaphors compared to human metaphors in terms of perceived enjoyment, intention to adopt, trust, likeability, and intelligence in the domains of health and finance? }
    \item \textbf{RQ2:} \textit{How does awareness of the metaphor (implicit versus explicit) impact these perceptions? }
\end{itemize}



 With this study, we aim to provide a starting point for conversation designers and researchers to explore the use of non-human metaphors in their design and help conceptualize a new design space focusing on non-human metaphors that might not act human but are human-centered. 

\section{Background and Related Work}\label{background}
\subsection{Metaphors: A Brief Overview}

Beyond literary devices used in poetry, metaphors can be interpreted as a source of sense-making~\cite{Cameron_Maslen_2010}. Each use of metaphors includes a source domain and a target domain, and there is some relationship between them ~\cite{Indurkhya_2013}. For example, in the phrase ``life is a journey'' life can be considered the target domain, and journey is the source domain. The concept of journey is explained using the existing understanding of life. Similarly, in the phrase ``time is money,'' the concept of money is understood using time. In both these cases, one concept is related to another, and the `relatively' more abstract concept is explained using a more concrete concept. These categories of metaphors are called \textit{conceptual metaphors}, and they gained prominence with the book ``Metaphors to Live By'' by cognitive linguists Lakoff and Johnson ~\cite{Lakoff_Johnson_1980}. 

Understanding conceptual metaphors goes beyond merely establishing relationships between familiar concepts; it also involves making sense of entirely new concepts by relating them to more familiar ones. Graphical User Interfaces (GUIs) exemplify this through their use of metaphors. GUIs rely on a primary `desktop' metaphor, which has undergone numerous iterations to provide users with a mental model for navigating visual interfaces~\cite{Newman_Sproull_1979}. Additionally, the primary desktop metaphor is complemented by various secondary metaphors (e.g., folders, files, bins) and auxiliary metaphors (e.g., scrolling, icons, logging) — we recommend Alan Blackwell's article~\cite{Blackwell_2006} for a comprehensive overview of the history of desktop metaphor and the idea of using metaphor as a design tool. These metaphors are so ubiquitous in the language of computing that we have ceased to notice them~\cite{Colburn_Shute_2008}, thereby becoming `dead' metaphors ~\cite{Reimer_1996}.

Comparatively, the history of metaphors in CUIs is nascent. CUIs are an umbrella term for a wide variety of technologies, including chatbots, Embodied Conversational Agents (ECAs), Virtual Assistants (VAs), and Voice User Interfaces (VUIs) or voice assistants (VAs). While our primary focus in this article is on VUIs, we also provide background on metaphors in CUIs to provide a more nuanced perspective on the current literature landscape. 

\subsection{Metaphors in CUIs}

 In ``Elements of Friendly Software Design''~\cite{Heckel_1984}, Paul Heckel described two fundamental functions of metaphor within interface design. The first function involves \textit{familiarization}, wherein metaphors are utilized to imbue the user interface with familiar concepts, enhancing comprehension and enabling users to construct mental models more easily. The second function, \textit{transportation}, revolves around unifying these mental models and facilitating immersive user experiences. Users are transported into a cohesive experiential realm by ingeniously supplementing familiar concepts to create novel experiences. In CUIs, the pervasive use of humanness as an operational metaphor by conversation designers serves as the vehicle for transportation. 

The humanness metaphor is realized through the creation of \textit{personas} (also known as `system personas'\footnote{\url{https://developers.google.com/assistant/conversation-design/create-a-persona}}) and \textit{social role reproduction} ~\cite{McMillan_Jaber_2021, Kim_Molina_Rheu_Zhan_Peng_2023}. These personas or social roles serve as the interface's front-end or interaction layer, shaping the user experience. Crafting personas is regarded as one of CUI design's most challenging and critical aspects~\cite{Sadek_Calvo_Mougenot_2023}. This involves scripting dialogues that reflect anticipated human behavior, choosing appropriate avatars for chatbots, or selecting voices that resonate with users' expectations. To this end, researchers and designers have deployed CUIs to fulfill various social roles, such as `teachers'~\cite{Desai_Chin_2023, Jung_Kim_So_Kim_Oh_2019}, `coaches'~\cite{Desai_Hu_Lundy_Chin_2023, Wang_Yang_Shao_Abdullah_Sundar_2020}, `storytellers'~\cite{Desai_Lundy_Chin_2023}, `brokers'~\cite{Lee_Frank_IJsselsteijn_2021} or `therapists'~\cite{Motalebi_Cho_Sundar_Abdullah_2019}. 

Designing metaphorical CUIs has an impact on how users perceive these interfaces. Khadpe et al.~\cite{Khadpe_Krishna_Fei-Fei_Hancock_Bernstein_2020} presented metaphorical chatbots with varying levels of competence and warmth (e.g., ``teenager,'' ``shrewd executive,'' or ``toddler'') using the prompt, ``The bot you are about to interact with is modeled after a \{ Metaphor \}.'' They found that the choice of metaphor had an impact on users' pre-use as well as post-use perceptions. Metaphors that signaled high competence were able to attract new users, but low competence metaphors were better evaluated and more likely to be adopted in the long term. Similarly, in another study, Chin et al. ~\cite{Chin2024Like} examined the impacts of formal and informal conversational styles of a VUI on the metaphorical descriptions provided by older and middle-aged adults. Their findings revealed that older adults tended to employ more professional metaphors (e.g., ``librarian,'' ``teacher,'' ``lawyer'') when describing formal VUIs while opting for personal metaphors (e.g., ``aunt,'' ``friend,'' ``child'') when describing informal VUIs. Researchers have also found relationships between the personalities projected by the metaphors and the personalities of the end-users. Braun et al. ~\cite{Braun_Mainz_Chadowitz_Pfleging_Alt_2019} designed VUI personalities projected by metaphors such as ``friend,'' ``admirer,'' ``aunt,'' and ``butler'' to understand user perceptions of in-car interactions. They found that users found metaphors that were similar to them (in terms of: openness, conscientiousness, extraversion, agreeableness, and neuroticism (OCEAN)~\cite{McCrae_Costa_1986}) were perceived to be more likable and trustworthy.  

While the ideal scenario involves presenting users with a clear mental model through personas, the reality is far more complicated due to the complex nature of users' understanding of conversational systems. Consider VUIs, also called `voice agents' or `voice assistants', or even Large Language Models (LLMs) like Microsoft's Copilot.\footnote{\url{https://copilot.microsoft.com}} In these cases, the persona is embedded in how we identify these technologies. However, despite this, the users often form their own conceptual understandings of these systems through folk theories~\cite{DeVito_Birnholtz_Hancock_French_Liu_2018}. Often, these folk theories are disseminated in the form of metaphors, including explanations for how CUIs work (e.g., ``\textit{it's like a computer chip that has stuff on it}'' ~\cite{Kim_Choudhury_2021} or ``\textit{genie in the bottle}'' ~\cite{Kuzminykh_Sun_Govindaraju_Avery_Lank_2020}) or making sense of conversational errors (e.g., ``\textit{like a silly child}''~\cite{Desai_Twidale_2022}). These folk theories stem from the discrepancy between what the users expect based on the metaphors provided to them and the actual experience of using CUIs~\cite{Luger_Sellen_2016}, also explained by the concept of Norman's ``gulf of execution''~\cite{Norman_2013}. This miscalibration can result in infrequent use~\cite{10.1145/3098279.3098539} or non-use~\cite{10.1145/3322276.3322332}. 

Moreover, operationalizing humanness as a metaphor raises ethical concerns, as highlighted by the CUI community in provocation papers\footnote{Provocation papers explore controversial, risk-taking, or nascent ideas that have the potential to spark debate.}. In CUI'21, McMillan \& Jaber~\cite{McMillan_Jaber_2021} discussed the inappropriateness of subordinate metaphors like ``assistants,'' ``maids,'' or ``servants.'' They argue that historically, these roles have been occupied by women of color (treated as ``physical absorbers'' of ``physical and affective dirt of a home''~\cite{Schiller_McMahon_2019}). Additionally, the presence of subordinate roles also implies the presence of ``master'' roles—a term loaded with unwanted connotations. Also, in CUI'21, Pradhan \& Lazar~\cite{Pradhan_Lazar_2021} raised the ethical issues with designing human-like personas, highlighting how the default voices of commercial VUIs often resemble ``white women from the West'' and emphasized that crafting female personas in subordinate roles can perpetuate existing societal biases. In CUI'22, Simpson \& Crone~\cite{Simpson_Crone_2022} vividly portrayed the potential dystopian consequences of designing CUIs with personas embodying societal roles, such as police officers, doctors, and priests. Their analysis revealed alarming outcomes, including the erosion of trust and the persistent manipulation of user behavior. Similarly, in CUI'22 Desai \& Twidale ~\cite{Desai_Twidale_2022} argued that the current implementation of personas in design is simplistic and advocated for contextually fluid personas that can change their behavior based on the task the user is performing. These provocations highlight the need for a more refined approach to designing metaphorical CUIs that fosters familiarity \textit{and} avoids ethical dilemmas.

An alternative and innovative approach to circumvent the implementation challenges and ethical considerations associated with designing CUIs based on human metaphors is to leverage non-human metaphors. Desai \& Twidale~\cite{Desai_Twidale_2023} developed a framework of metaphor contextualization in VUIs by identifying metaphors used by users in semi-structured interviews and VUI literature. They found that users did employ a broad range of human metaphors to describe VUIs, \textit{but} they also used non-human metaphors (e.g., tower, ``Computer'' from Star Trek, genie, encyclopedia). Moreover, users fluctuated between human and non-human metaphors seamlessly without cognitive dissonance. This finding is also supported by Pradhan et al.~\cite{Pradhan_Findlater_Lazar_2019}, who found older adults ontologically categorized VUIs as a person as well as an object (or something in between). In another relevant study, Jung et al. ~\cite{Jung_Qiu_Bozzon_Gadiraju_2022} used the Great Chain of Being framework (GCOB)\footnote{Theorized hierarchical categorization of the world using the metaphors of god, human, animal, plant, and an inorganic object.}~\cite{Jackendoff_Aaron_1991} to understand the effect of non-human metaphors (God, Animal, Plant, and Book) on user perceptions compared to a human metaphor, when applied to chatbots. They found that the human metaphor had no significant difference in chatbot evaluation over non-human metaphors. In the current study, we build on (1) Desai \& Twidale's~\cite{Desai_Twidale_2023} work on metaphor contextualization in VUIs, and (2) further explore Jung et al.'s~\cite{Jung_Qiu_Bozzon_Gadiraju_2022} findings on designing non-human metaphors in the context of VUIs. 

\subsection{Application of VUIs}
Health and finance are two popular application areas of VUIs~\cite{alnefaie2021overview}. In our investigation, we focus on these two areas since they can be considered as essential for individuals' well-being and thus we believe that the presented interaction contexts will be familiar and relatable to most VUI users. Moreover, due to the sensitive nature of heath and finance, trust will play an important role in the voice interactions~\cite{ammari2019music}.  Below, we discuss the characteristics of both of these domains.

\subsubsection{Applications in Health}
Healthcare is one of the most prolific sectors for VUI deployment~\cite{lupetti2023trustworthy}. Applications of VUIs in this area include: screening and monitoring~\cite{kocielnik2021can,halperin2023probing}, consulting and advice~\cite{van2017potential,lupetti2023trustworthy}, learning~\cite{Desai_Chin_2023}, mental health support~\cite{vaidyam2019chatbots,miner2016smartphone}, or healthy lifestyle advice~\cite{fadhil2019assistive,cheng2018development}. 

Some most commonly used personalities in this sensitive setting are coach-like, healthcare-expert, professional-like, informal, etc.~\cite{tudor2020conversational}.
While users' preferences regarding desired VUI personality may vary~\cite{volkel2021manipulating}, it has been demonstrated that the type of VUI personality can directly impact users' attitudes and interactions~\cite{lupetti2023trustworthy}. For instance, conversational style choices such as using humor, providing personalized and timely feedback, or starting the conversation with a social chat have proven beneficial to enhancing the overall user experience~\cite{bickmore2009taking}. Cox et al. \cite{cox2022does} found that formal conversation style was perceived positively when conversation concerned managing sensitive health information or discussing patient's lifestyle.

Subject sensitivity and privacy concerns are important considerations that can affect users' intention to disclose information regarding their health online~\cite{bansal2010impact}. However, it should be noted that same subject may have different levels of sensitivity for different different people or even for the same person depending on the circumstances ~\cite{schuetzler2018influence}. Schuetzler et al. shown that even small differences in communication style such as increasing responsiveness can have significant impact on the quality of information gathered during the interview ~\cite{schuetzler2018influence}. This is particularly important given that VUIs are perceived as free from personal bias and induce less anxiety in patients when disclosing risky health behaviors~\cite{luxton2020ethical}.

\subsubsection{Applications in Finance}
When it comes to finance, VUIs have been adopted for tasks such as providing customer support or enabling users to execute certain orders and transactions \cite{reicherts2022extending}. In the investing context, conversational robo-advisors are a popular use case ~\cite{candello2017shaping,day2018artificial,hildebrand2021conversational,morana2020effect}. Usually, the goal of such advisors is to capture user's interests and preferences and probe their risk aversion, in order to provide suggestions for suitable financial products, investment strategies, and/or portfolio allocations \cite{reicherts2022extending}. By providing individualized investment suggestions, robo-advisors not only facilitate the on-boarding experience for the users, but also help service provider to streamline their customer support process and improve customer retention. Reicherts et al. have developed a chatbot to probe and scaffold decision-making process of investors, maintain their investment strategy and avoid emotional reactions \cite{reicherts2022extending}. The proposed system was successful in promoting self-reflection and user's awareness of their thought processes that are fundamental for strategic thinking and decision making.

Hodge, Mendoza, and Sinha~\cite{hodge2021effect} examined the effect of anthropomorphization of robo-advisors in the context of investor judgments by manipulating the type of advisor (human vs. robo-advisor) and the level of humanization (low vs. high). They found that investors were more likely to follow recommendations from a robo-advisor with a low-level of humanization. On the other hand, recommendations from human advisors exhibiting a high level of human characteristics were more likely to be followed. Related studies similarly indicated that anthropomorphizing VUIs can increase their perceived social presence and, in turn, affects users’
propensity to trust them~\cite{araujo2018living,qiu2009evaluating}. Recently, Morana et al. demonstrated that appropriate design of social clues design in finance robo-advisors is essential to achieve adequate level of social presence and, in turn, increase user's trust in the agent~\cite{morana2020effect}.


\section{Study}\label{methodology}
The study employed a 2x4 factorial design to investigate the effects of metaphor type (human vs. non-human) and awareness level (explicit vs. implicit) on user perceptions of VUIs in health and finance domains. Within this design, two independent variables were manipulated: Awareness Level, with two between-subjects levels (Explicit Awareness, Implicit Awareness), and Metaphor Type, with four within-subjects levels (Doctor, Health Encyclopedia in Health; Financial advisor, Calculator in Finance). This approach allows for examining the main effects and interactions between different types of metaphors and the user's awareness regarding the metaphor's presence. Participants experienced all four metaphor types within their assigned awareness level, facilitating a within-subjects comparison for metaphor types and a between-subjects comparison for awareness levels. This between-groups design facilitates disentangling the individual and combined influence of metaphor type and awareness on perceived enjoyment, intention to adopt, trust, intelligence, and likeability in VUI interactions.
\subsection{Designing Metaphorical VUIs}
\subsubsection{Metaphor Selection}
We selected four metaphors from a database compiled by Desai \& Twidale ~\cite{Desai_Twidale_2023}, which catalogs metaphors employed by users or designers to articulate their interactions with VUIs. Jung et al.'s investigation ~\cite{Jung_Qiu_Bozzon_Gadiraju_2022} into chat-based conversational interfaces, employing the GCOB framework, revealed that the inorganic object metaphor notably diverged in perception from the human metaphor. This prompted us to explore whether such distinctions would also be evident in the context of VUIs, leading us to juxtapose a human metaphor against a domain-specific (inorganic object) non-human metaphor. Further, drawing on findings from prior research \cite{Jung_Qiu_Bozzon_Gadiraju_2022, Khadpe_Krishna_Fei-Fei_Hancock_Bernstein_2020} that users more favorably receive metaphors signaling high competence and warmth, we opted for \textbf{\textit{`doctor'}} as the human metaphor and \textbf{\textit{`health encyclopedia'}} as the non-human metaphor within the health domain. Similarly, for finance, we chose \textbf{\textit{`financial advisor'}} and \textbf{\textit{`calculator'}} as the human and non-human metaphors, respectively.
\subsubsection{Conversation Design}
Our study involved conversations between users and metaphorical VUIs created by the first two authors, each with eight years of expertise in conversation design. These conversations were structured to reflect how `Z,' in human and non-human versions, would communicate identical information across different domains. The design intricacies of these dialogues are presented in Table \ref{table 1}, which showcases snippets of these interactions, allowing for a direct comparison between the delivery styles of the human and non-human versions of `Z.'
Each script, encompassing human and non-human `Z' in the health and finance domain, consisted of six turns, with the user prompts remaining consistent across both domains. The variable element was the responses from `Z,' uniquely tailored based on the metaphorical framework employed to convey the information. This approach of designing metaphor-driven scripts aligns with established conversation design methodologies~\cite{Sadek_Calvo_Mougenot_2023}, wherein designers meticulously craft script-level templates that embody each chosen metaphor. To ensure the metaphors effectively shaped the user's experience, our design process was anchored in several key considerations, instrumental in differentiating the communication styles of `Z,' including (1)~non-human metaphors took a more direct and didactic approach to answering user prompts, (2)~human metaphors were designed to be more informal and conversational, and (3)~non-human metaphors cited sources when providing answers. 
\subsubsection{Synthesizing Voices}
The audio clips used in our study were generated through a text-to-speech (TTS) software, Speechify.\footnote{\url{https://speechify.com/}} This selection was motivated by Speechify's capability to simulate conversations using multiple synthetic voices, thereby enhancing the realism of the audio clips. Initially, our design strategy involved employing a human voice actor for the user's role, with the TTS software being utilized solely for the voice of the VUI. However, during preliminary testing, we encountered challenges maintaining consistent pitch and tone with the human voice actor across the different experimental conditions. Given that prior research, such as Dubiel et al.~\cite{dubiel2024impact,dubiel2020persuasive}, has indicated that even minor variations in voice characteristics can significantly influence user perceptions, we decided to adopt a synthetic voice for the user role as well. In selecting the synthetic voices for both the VUI and the user, we considered the prevalent gender representation in commercial VUIs, such as Siri and Alexa, which typically feature female-sounding voices~\cite{curry2020conversational}. To establish a clear distinction between the VUI and the user voices, we opted for a female-sounding voice for the VUI and a male-sounding voice for the user. This choice was also informed by literature indicating the potential impact of voice gender on user perception~\cite{Kuzminykh_Sun_Govindaraju_Avery_Lank_2020}. Furthermore, in an effort to mitigate any potential biases linked to the VUI's identity, we named the VUI `Z'. This name was chosen for its neutrality, lacking associations with any particular gender or race, thus aligning with our objective to minimize confounding variables that could influence the study's outcomes. To maintain consistency across all four scripts, we used the male synthetic voice named ``Guy'' for the user and set the speaking tone to ``chat'' to make the voice sound more human-like. For `Z,' we used the female voice ``Aria'' and set the speaking tone to ``none''. The speech, speed, and volume modifications were all set to zero percent. For the explicit awareness condition, the conversation was preceded by the prompt ``You are about to hear a conversation between a user with a male-sounding voice and a voice interface named `Z', with a female-sounding voice. The voice interface `Z' is modeled after a \{ Metaphor \} in this conversation. The conversation will start now.'' For implicit awareness, we removed the ``The voice interface `Z' is modeled after a \{ Metaphor \} in this conversation'' part of the prompt. This prompt is adapted from~\cite{Khadpe_Krishna_Fei-Fei_Hancock_Bernstein_2020}. All audio files are included as supplementary material. 

\begin{table}[]
\caption{Snippets of responses provided by human and non-human versions of `Z' in health and finance domains.}
\label{table 1}
\begin{tabular}{l|p{2.5in}|p{2.5in}}
        & \multicolumn{1}{c|}{\textbf{Human}} & \multicolumn{1}{c}{\textbf{Non-Human}} \\ \toprule
\textbf{Health}  & \textbf{\textit{User: }} Hey `Z,' I have heard that cardio exercise is good for you, but I'm not sure why. Can you explain the benefits.
\newline\newline\textbf{\textit{Doctor:}} Of course! Cardio is great for your overall health. It gets your heart rate up, which strengthens your heart and improves your cardiovascular system.
\newline\newline\textbf{\textit{User:}}  Okay, and how often should I do cardio?
\newline\newline\textbf{\textit{Doctor:}} Aim for at least 150 minutes of moderate-intensity or 75 minutes of high-intensity cardio per week. But it is not all or nothing—every bit helps. Even if you can't meet these guidelines, some exercise is better than none. And remember, it's important to find activities you enjoy; it makes it much easier to stick with it.& 
\textbf{\textit{User: }} Hey `Z,' I have heard that cardio exercise is good for you, but I'm not sure why. Can you explain the benefits.
\newline\newline\textbf{\textit{Health Encyclopedia:}} Cardiovascular exercise, often referred to as cardio, encompasses any rhythmic activity that raises your heart rate into your target heart rate zone. 
\newline\newline\textbf{\textit{User:}}  Okay, and how often should I do cardio?
\newline\newline\textbf{\textit{Health Encyclopedia:}} The American Heart Association recommends at least 150 minutes of moderate-intensity aerobic activity or 75 minutes of vigorous aerobic activity per week, ideally spread throughout the week. Even short bouts of activity, like 10 minutes, can be beneficial if performed regularly.          \\ \midrule
\textbf{Finance} & \textbf{\textit{User: }} Hi `Z,' I'm trying to figure out how much life insurance I might need. Can you help me with that? 
\newline\newline\textbf{\textit{Financial Advisor: }} Great, a common starting point is to aim for 10 to 15 times your annual salary in life insurance coverage. However, this is a simplistic rule and doesn't account for all variables. Let's delve deeper. Do you have any significant debts like a mortgage or car loans? 
\newline\newline\textbf{\textit{User:}}  That's a bit more than I expected. Is there a way to reduce the cost?
\newline\newline\textbf{\textit{Financial Advisor: }} To manage costs, you might consider term life insurance. It's typically less expensive than whole life insurance and can be structured to cover your most financially vulnerable years, like until your mortgage is paid off or your children are through college. Remember, the cheapest policy isn't always the best. It's about finding the right balance between coverage and affordability. & \textbf{\textit{User: }}   Hi `Z,' I'm trying to figure out how much life insurance I might need. Can you help me with that? 
\newline\newline \textbf{\textit{Calculator:}} Based on your income, a general estimate for your life insurance coverage would be between \$500,000 (which is 10 times your income) and \$750,000 (which is 15 times your income). Now, let's consider any outstanding debts you have. Do you have any debts like a mortgage or car loans?
\newline\newline\textbf{\textit{User:}}  That's a bit more than I expected. Is there a way to reduce the cost?
\newline\newline \textbf{\textit{Calculator:}} One way to manage the cost is to adjust the term or type of insurance. For example, term life insurance is generally less expensive than whole life insurance and provides coverage for a set period. You can also adjust the coverage amount based on your most critical financial needs to find a balance between sufficient coverage and affordability.
\end{tabular}
\end{table}

\subsection{Measures}

To obtain a comprehensive insight into users' perceptions of metaphorical VUIs, we adopt the classification proposed by Wei et al.~\cite{Wei_Kim_Kuzminykh_2023}. This classification categorizes user perceptions of CUIs into two relevant categories: (1) perception of interaction with agents and (2) perception of agent's characteristics. The classification is derived from a literature review encompassing measures commonly employed in CUI research to evaluate users' interactions with VUIs and their perceptions of the agents' characteristics. Guided by this framework, our study will incorporate the following measures:
\begin{itemize}
\item \textit{\textbf{Perception of interaction with agents}} encompasses evaluative metrics to gauge the overall quality of interaction between a CUI and a user. This facet includes elements such as engagement, focusing on constructs like perceived enjoyment and intention to adopt. Within VUI literature, there is significant interest beyond mere usability, exploring how VUIs can contribute to user enjoyment~\cite{Yang_Aurisicchio_Baxter_2019} or enhance the quality of life ~\cite{Desai_Twidale_2023}. To assess \textit{\textbf{perceived enjoyment}}, we employ Moussawi et al.’s adapted survey~\cite{Moussawi_Koufaris_Benbunan-Fich_2021}, comprising three items on a 7-point Likert scale: ``While using Z, I would find the interaction enjoyable,'' ``While using Z, I would find this interaction interesting,'' and ``While using Z, I would find the interaction fun.'' To measure users’ \textbf{\textit{perceived intention to adopt}}, we adapt Moussawi et al.’s survey~\cite{Moussawi_Koufaris_Benbunan-Fich_2021}, incorporating two items on 7-point Likert scale: ``If available, I intend to start using Z within the next month,'' and ``If available, in the next months, I plan to experiment or regularly use Z.'' Both these measures were developed and adapted to be used for VUIs. 
\item \textbf{\textit{Perception of agent's characteristics}} comprises metrics designed to elucidate how users perceive a CUI's capabilities and personality traits, including competence, likeability, and trustworthiness. To examine variations in perceptions of the VUI’s characteristics across different conditions, we utilized \textbf{\textit{perceived intelligence}} and \textbf{\textit{likeability}} measures from the Godspeed Measures using 5-point semantic differential scales ~\cite{Bartneck_2023, Bartneck_Kulić_Croft_Zoghbi_2009}, which are widely employed in VUI research~\cite{Wei_Kim_Kuzminykh_2023, Seaborn_Urakami_2021}. Additionally, \textbf{\textit{perceived trust}} was assessed using a 7-point Likert scale adapted from Jian et al.~\cite{Jian_Bisantz_Drury_2000}.
\end{itemize}
Further, we allowed participants the option to provide qualitative feedback regarding their interactions for each version of `Z,' should they choose to do so.

\subsection{Procedure}
Participants for this study were recruited via Prolific\footnote{\url{https://www.prolific.com/}}, an online research platform connecting researchers with a large participant pool. They were compensated \$3.0 each for approximately 20 minutes of their time, reflecting the federal minimum wage in the U.S. The process began with participants accessing the study through a link provided by Prolific, where they first encountered and consented to an online consent form—as mandated by policies of the Institutional Review Board (IRB). Subsequently, they completed a demographic questionnaire to gather basic information like age, gender, educational background, and prior experience and familiarity with VUIs.

Following demographic data collection, participants were randomly divided into two groups based on their level of awareness—Explicit Awareness or Implicit Awareness—in a between-subjects design framework. They then partook in a 2x4 factorial design experiment, interacting with four VUI metaphor types across two domains, Health and Finance, in a within-subjects approach. In the Health domain, interactions involved a `Doctor' (Human metaphor) and a `Health Encyclopedia' (Non-Human metaphor), whereas, in the Finance domain, they engaged with a `Financial Advisor' (Human metaphor) and a `Calculator' (Non-Human metaphor). Audio clips simulating these interactions, generated via the text-to-speech service Speechify, provided a controlled and uniform experience for all participants. The sequence of metaphor presentation was counterbalanced to mitigate order effects. 
After each audio clip, participants completed questionnaires designed to evaluate their perceptions of the VUIs, capturing metrics such as perceived enjoyment, intention to adopt, likeability, intelligence, and trust. Additionally, they were given the opportunity to express their subjective views on the audio clips through an optional text box for open-ended feedback. To ensure attentive participation, an attention-check audio clip was also introduced; following this clip, participants were required to select specific responses in the subsequent questionnaires, helping to validate the integrity of their engagement throughout the study.

Upon completion of interactions with all four metaphor types across the two domains, participants received a Prolific-generated completion code. This code served to verify their participation on the Prolific platform and to ensure they received their compensation, effectively concluding the study procedures.

\subsection{Participants}

Our study design accounted for statistical power in the planning stages to ensure the reliability of the results. Guided by a power analysis, we determined that a total sample size of at least 232 participants would be necessary to observe a difference of one standard deviation between conditions with our 2x4 factorial design, assuming an alpha level of 0.05 (to minimize Type I errors) and a power level of 0.80 (to minimize Type II errors). Participants were stratified across our experimental conditions to maintain balance. Moreover, since our study required participants to imagine themselves using `Z,' we required them to have prior experiences with voice interfaces. Additionally, the use and understanding of metaphors are based on various cultural factors~\cite{Moser_2000}, so we limited our study to highly proficient English speakers based in the U.S.

We recruited 240 participants (M=43.21, SD=13.73) via Prolific with a minimum task approval rate of 98\% and at least one year of activity on the platform. Among the recruited participants, 49.2\% identified as female, 49.6\% as male, 0.8\% as non-binary, and 0.4\% preferred not to disclose their gender. Additionally, 78.3\% of participants held a Bachelor's degree or higher. All participants resided in the U.S. and either spoke English as their native language or used it as their primary language of communication. Furthermore, 82.5\% of participants reported being very or extremely familiar with VUIs. Every participant used VUIs such as Alexa or Siri on a weekly basis, with 58.3\% using them daily. Participants received compensation at an hourly rate of \$8.33/hr, aligning with federal minimum wage regulations in the U.S.

\section{Results}\label{results}

In assessing the impact of metaphor type (human vs. non-human) across health, finance, and overall, a repeated measures ANOVA was utilized. Metaphor type served as the within-subjects variable, while participant awareness (implicit vs. explicit) was the between-subjects factor. This design aimed to isolate the effects of metaphor type and awareness on perceived enjoyment, intention to adopt, trust, intelligence, and likeability. 

\subsection{Perceived Enjoyment}
\begin{figure}[h!]
  \centering
  \includegraphics[width=1\textwidth]{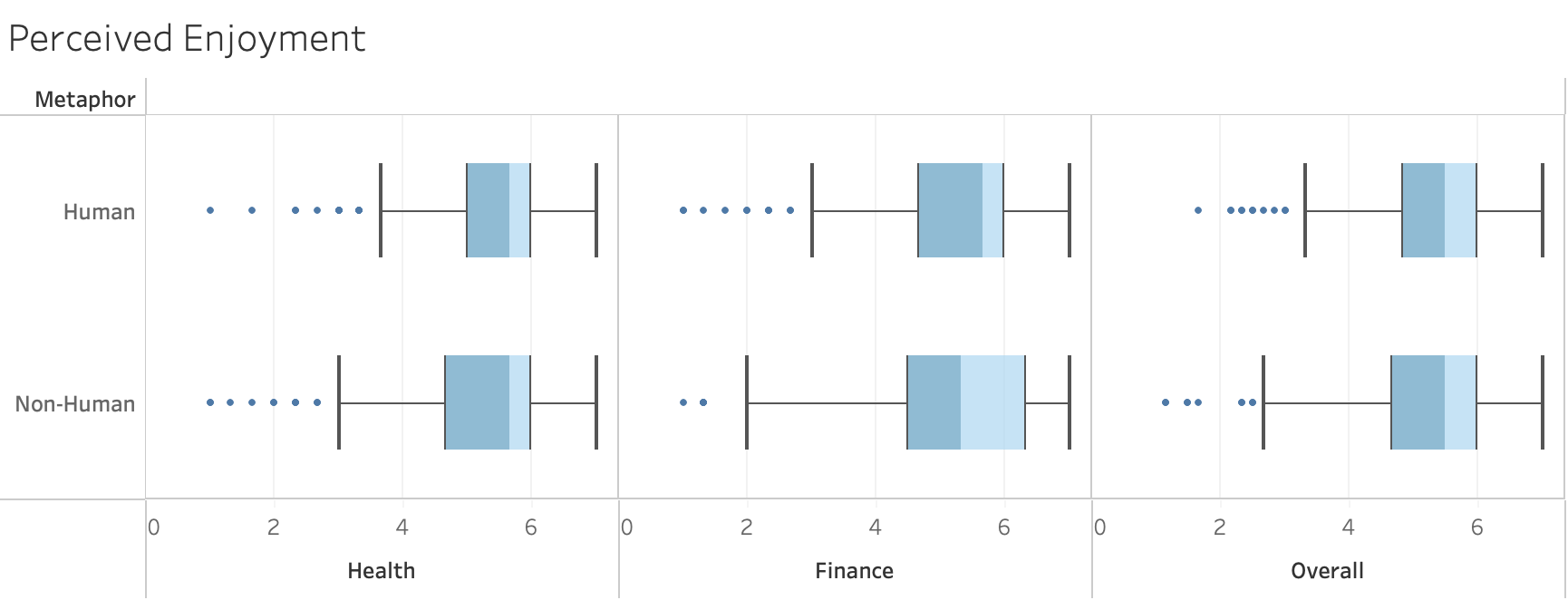}
  \caption{Boxplots indicating the distribution of perceived enjoyment scores for human and non-human metaphors across health, finance, and overall.}
  \label{fig:1}
\end{figure}

For health, the analysis revealed a significant main effect of metaphor type on perceived enjoyment ($F(1, 238) = 4.663, p = .032, \eta_{p}^{2} = 0.019$). This indicates that the type of metaphor had a significant impact on participants' enjoyment, with the human metaphor leading to higher perceived enjoyment than the non-human metaphor. No significant interaction effect between metaphor type and awareness level on perceived enjoyment was found ($F(1, 238) = 0.376, p = .540, \eta_{p}^{2} = 0.002$), suggesting that the impact of metaphor type on perceived enjoyment was not significantly different across levels of awareness. The interaction between metaphor type and awareness levels did not significantly affect perceived enjoyment in the finance domain ($F(1, 238) = 0.045, p = .832, \eta_{p}^{2} = 0.000$). No significant interaction effect of the awareness level on metaphor type was found as well ($F(1, 238) = 0.023, p = .879, \eta_{p}^{2} = 0.000$). Overall, over both domains (health and finance), we found that there was a significant main effect of metaphor type on perceived enjoyment ($F(1, 238) = 9.798, p = .002, \eta_{p}^{2} = 0.040$). However, again, there was no interaction effect of awareness on perceived enjoyment ($F(1, 238) = 1.270, p = .261, \eta_{p}^{2} = 0.005$). The means and standard deviations of all metaphors across each domain are specified in \autoref{table 2}. \autoref{fig:1} shows the distribution of perceived enjoyment scores.

To summarize, we found that participants perceived interacting with the human metaphor to be more enjoyable than the non-human metaphor for the domains of health and overall. No significant differences were found in the perceived enjoyment of human and non-human metaphors in finance. 

\begin{table}[h!]
\caption{The perceived enjoyment scores (µ ± $\sigma$: mean and standard deviation) of human and non-human metaphors for the domains of health, finance, and overall.}
\label{table 2}
\begin{tabular}{llll}
\toprule
\multicolumn{4}{c}{\textbf{Perceived Enjoyment}} \\ \midrule
 & \multicolumn{1}{c}{Health} & \multicolumn{1}{c}{Finance} & \multicolumn{1}{c}{Overall} \\
Human           & $\qty{5.43 \pm 1.12}{\percent}$       & $\qty{5.27 \pm 1.29}{\percent}$        & $\qty{5.35 \pm 1.13}{\percent}$      \\
Non-human       & $\qty{5.32 \pm 1.24}{\percent}$      & $\qty{5.26 \pm 1.30}{\percent}$        & $\qty{5.27 \pm 1.25}{\percent}$      \\ \bottomrule
\end{tabular}
\end{table}

\subsection{Perceived Intention to Adopt}
\begin{figure}[h!]
  \centering
  \includegraphics[width=1\textwidth]{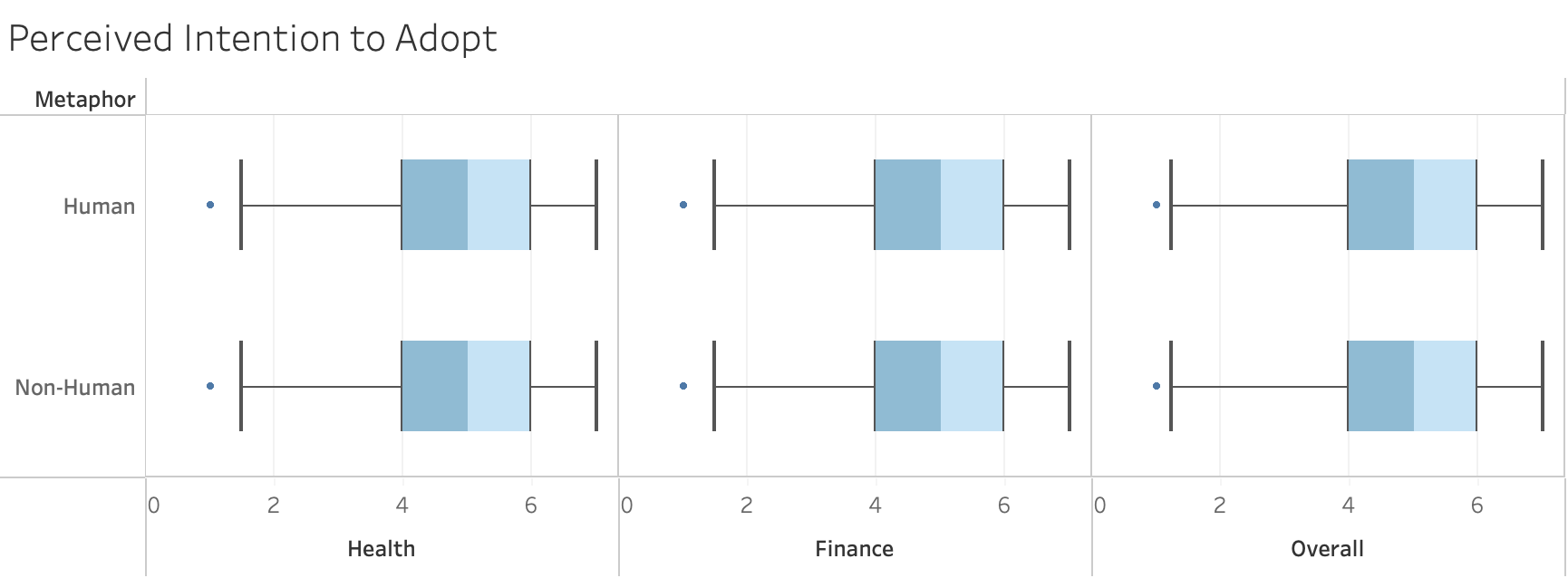}
  \caption{Boxplots indicating the distribution of perceived intention to adopt scores for human and non-human metaphors across health, finance, and overall.}
  \label{fig:2}
\end{figure}

In the health domain, we found no significant main effect of metaphor type on perceived intention to adopt ($F(1, 238) = 1.322, p = .251, \eta_{p}^{2} = 0.006$). Further, no significant interaction effect between metaphor type and awareness level on intention to adopt was found ($F(1, 238) = 0.266, p = .607, \eta_{p}^{2} = 0.001$). Similarly, the interaction between the human and the non-human metaphor and awareness did not significantly affect perceived intention to adopt in the finance domain ($F(1, 238) = 1.266, p = .262, \eta_{p}^{2} = .005$). We also did not find significant interaction effect of the awareness level on metaphor type ($F(1, 238) = 0.749, p = .388, \eta_{p}^{2} = 0.003$). Over both domains, we found no significant main effect of metaphor type on perceived intention to adopt ($F(1, 238) = 1.193, p = .276, \eta_{p}^{2} = 0.005$). Intriguingly, while overall metaphor type did not lead to differences in perceived intention to adopt, the significant interaction effect of awareness level suggests that within the conditions where non-human metaphors were utilized, the awareness level may play a more pronounced role in influencing intention to adopt ($F(1, 238) = 4.116, p = .044, \eta_{p}^{2} = 0.017$). \autoref{table 3} presents means and standard deviations, for each category of metaphors within their specific domains. \autoref{fig:2} indicates the distribution of the perceived intention to adopt scores. 

In sum, our findings suggest that human and non-human metaphors did not differ significantly in their impact on perceived intention to adopt across health and finance domains. However, interaction involving awareness level and non-human metaphors emerged as a significant factor in the overall intention to adopt.

\begin{table}[h!]
\caption{The perceived intention to adopt scores (µ ± $\sigma$: mean and standard deviation) of human and non-human metaphors for the domains of health, finance, and overall.}
\label{table 3}
\begin{tabular}{llll}
\toprule
\multicolumn{4}{c}{\textbf{Perceived Intention to Adopt}} \\ \midrule
 & \multicolumn{1}{c}{Health} & \multicolumn{1}{c}{Finance} & \multicolumn{1}{c}{Overall} \\
Human           & $\qty{4.92 \pm 1.44}{\percent}$       & $\qty{4.91 \pm 1.41}{\percent}$        & $\qty{4.91 \pm 1.37}{\percent}$      \\
Non-human       & $\qty{4.86 \pm 1.41}{\percent}$      & $\qty{4.86 \pm 1.45}{\percent}$        & $\qty{4.88 \pm 1.38}{\percent}$      \\ \bottomrule
\end{tabular}
\end{table}

\subsection{Perceived Trust}

\begin{figure}[h!]
  \centering
  \includegraphics[width=1\textwidth]{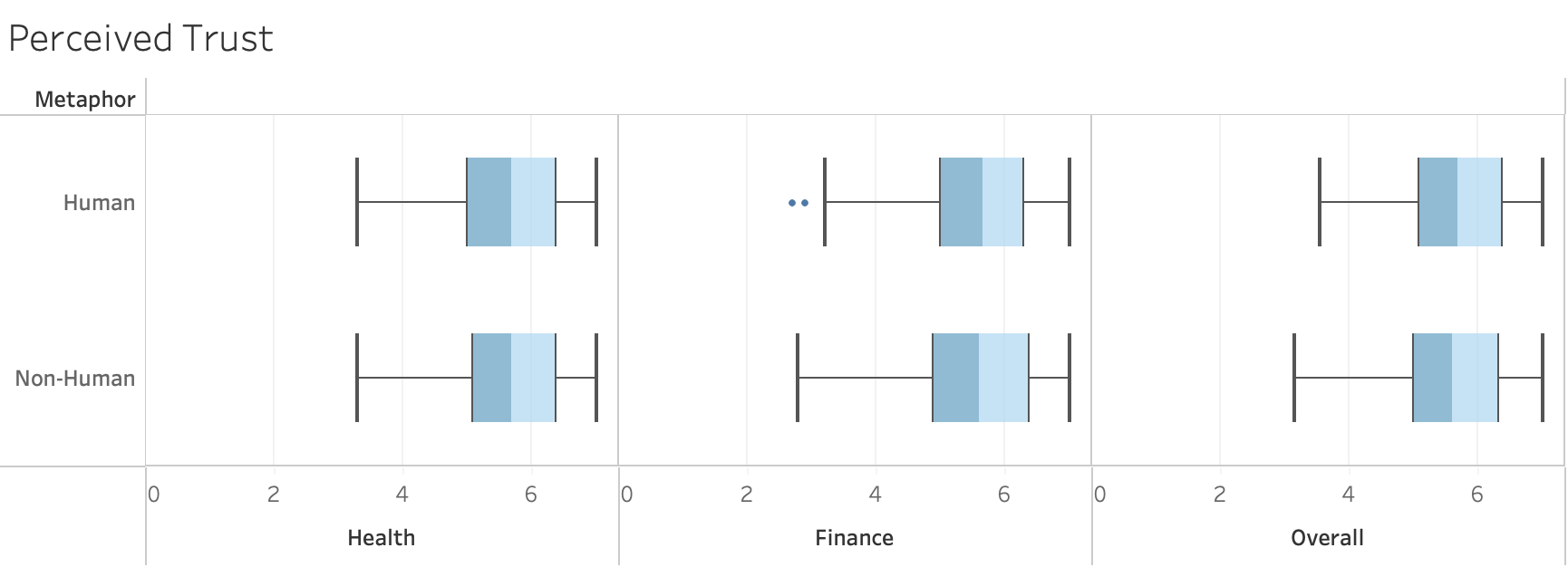}
  \caption{Boxplots indicating the distribution of trust scores for human and non-human metaphors across health, finance, and overall.}
  \label{fig:3}
\end{figure}

For health, our analysis found no significant main effect of metaphor type on perceived trust ($F(1, 238) = 0.796, p = .373, \eta_{p}^{2} = 0.003$). Additionally, we did not find a significant interaction effect between metaphor type and awareness level on perceived trust ($F(1, 238) = 0.017, p = .896, \eta_{p}^{2} = 0.001$). There was no main effect between metaphor type and awareness of trust in the finance domain ($F(1, 238) = 0.195, p = .659, \eta_{p}^{2} = 0.001$). Further, there was no significant interaction effect of the awareness level on metaphor type ($F(1, 238) = 1.349, p = .247, \eta_{p}^{2} = 0.006$). Interestingly, over the two domains, we found a significant main effect of metaphor type on trust ($F(1, 238) = 5.577, p = .019, \eta_{p}^{2} = 0.023$), with participants trusting the human metaphors more than non-human metaphors. However, there was also no interaction effect of awareness level ($F(1, 238) = 0.242, p = .624, \eta_{p}^{2} = 0.001$) on trust scores. \autoref{table 4} shows means and standard deviations for human and non-human metaphors over the domains. The distribution of trust scores can be seen in \autoref{fig:3}. 

\begin{table}[h!]
\caption{The perceived trust scores (µ ± $\sigma$: mean and standard deviation) of human and non-human metaphors for the domains of health, finance, and overall.}
\label{table 4}
\begin{tabular}{llll}
\toprule
\multicolumn{4}{c}{\textbf{Perceived Trust}} \\ \midrule
 & \multicolumn{1}{c}{Health} & \multicolumn{1}{c}{Finance} & \multicolumn{1}{c}{Overall} \\
Human           & $\qty{5.66 \pm 0.90}{\percent}$       & $\qty{5.59 \pm 0.95}{\percent}$        & $\qty{5.62 \pm 0.88}{\percent}$      \\
Non-human       & $\qty{5.63 \pm 0.90}{\percent}$      & $\qty{5.57 \pm 0.95}{\percent}$        & $\qty{5.58 \pm 0.91}{\percent}$      \\ \bottomrule
\end{tabular}
\end{table}

In summary, our results show that for health and finance, there were no significant differences in how participants perceived trusting human and non-human metaphors. However, overall, participants trusted human metaphors more than non-human metaphors. 

\subsection{Perceived Likeability}
\begin{figure}[h!]
  \centering
  \includegraphics[width=1\textwidth]{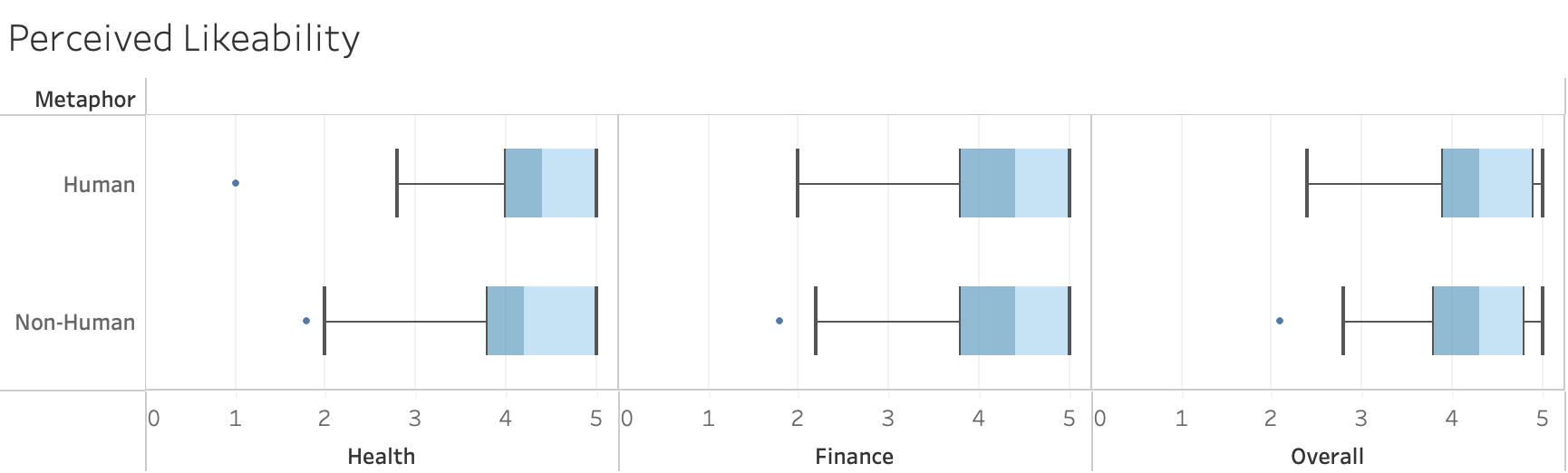}
  \caption{Boxplots indicating the distribution of likeability scores for human and non-human metaphors across health, finance, and overall.}
  \label{fig:4}
\end{figure}

We found a significant main effect of human and non-human metaphor on perceived likeability ($F(1, 238) = 7.551, p = .006, \eta_{p}^{2} = 0.031$), with participants considering human metaphor more likeable than the non-human metaphor. There was no significant interaction effect between metaphor type and awareness level on perceived likeability ($F(1, 238) = 0.043, p = .836, \eta_{p}^{2} = 0.000$), indicating no impact of metaphor type on likeability across explicit and implicit awareness levels. The interaction between metaphor type and awareness levels did not significantly affect perceived likeability finance ($F(1, 238) = 0.971, p = .325, \eta_{p}^{2} = 0.004$). There was also no significant interaction effect of the awareness of awareness level on metaphor type ($F(1, 238) = 1.333, p = .249, \eta_{p}^{2} = 0.006$). Over the two domains, we found a significant main effect of metaphor type on likeability ($F(1, 238) = 8.913, p = .003, \eta_{p}^{2} = 0.036$). However, there was no interaction effect of awareness on likeability ($F(1, 238) = 0.160, p = .690, \eta_{p}^{2} = 0.001$). The means and standard deviations of all metaphors across each domain are specified in \autoref{table 5} and \autoref{fig:4} show the distribution of perceived likeability scores.

To summarize, in the health domain, we found that participants perceived the human metaphor to be more likeable. There was no significant difference of metaphor type on likeability in finance. Overall, human metaphors were perceived to be more likeable. Awareness of metaphors did not influence likeability across all domains. 

\begin{table}[h!]
\caption{The perceived likeability scores (µ ± $\sigma$: mean and standard deviation) of human and non-human metaphors for the domains of health, finance, and overall.}
\label{table 5}
\begin{tabular}{llll}
\toprule
\multicolumn{4}{c}{\textbf{Perceived Likeability}} \\ \midrule
 & \multicolumn{1}{c}{Health} & \multicolumn{1}{c}{Finance} & \multicolumn{1}{c}{Overall} \\
Human           & $\qty{4.32 \pm 0.61}{\percent}$       & $\qty{4.25 \pm 0.69}{\percent}$        & $\qty{4.28 \pm 0.60}{\percent}$      \\
Non-human       & $\qty{4.22 \pm 0.69}{\percent}$      & $\qty{4.21 \pm 0.69}{\percent}$        & $\qty{4.23 \pm 0.63}{\percent}$      \\ \bottomrule
\end{tabular}
\end{table}

\subsection{Perceived Intelligence}
\begin{figure}[h!]
  \centering
  \includegraphics[width=1\textwidth]{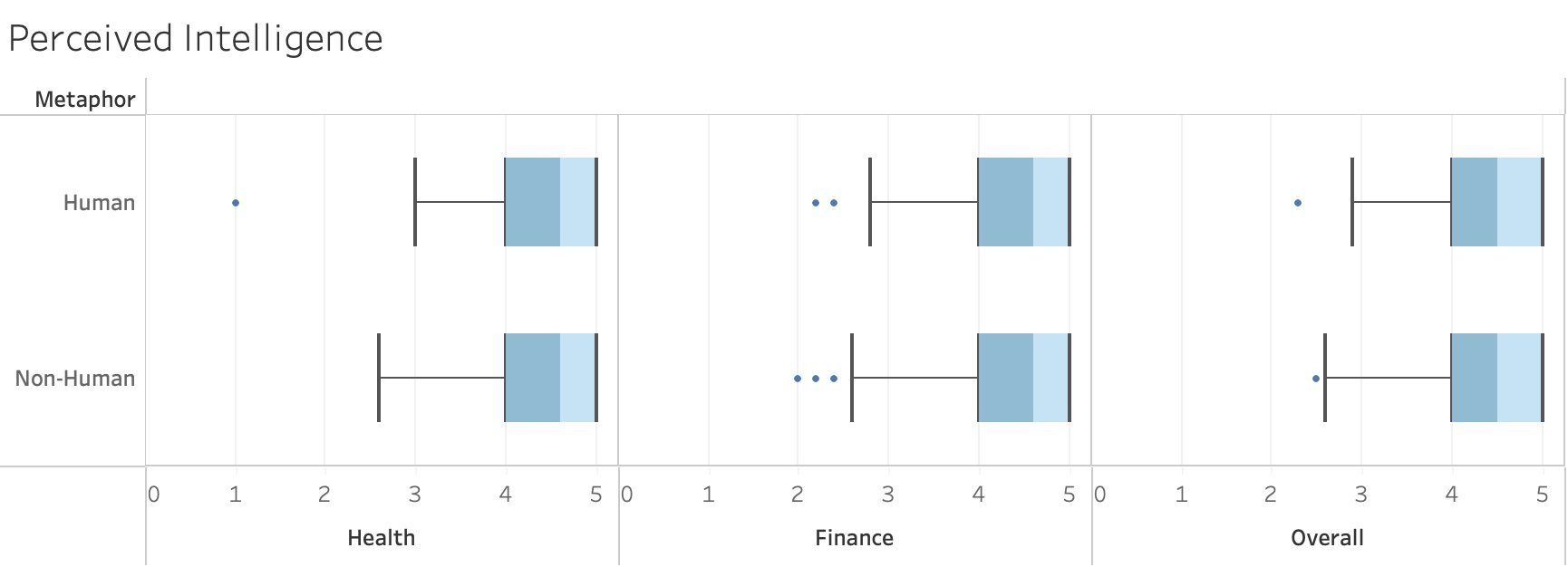}
  \caption{Boxplots indicating the distribution of perceived intelligence scores for human and non-human metaphors across health, finance, and overall.}
  \label{fig:5}
\end{figure}

In the context of health, there was no significant main effect of human and non-human metaphor on perceived intelligence ($F(1, 238) = 0.053, p = .818, \eta_{p}^{2} = 0.000$). There was also no significant interaction effect between metaphor type and awareness level on intelligence ($F(1, 238) = 0.030, p= .863, \eta_{p}^{2} = 0.000$). Similarly, there was no impact of metaphor type on perceived intelligence in the finance domain ($F(1, 238) = 0.971, p = .325, \eta_{p}^{2} = 0.004$). We also did not find a significant interaction effect of the awareness level on human and non-human metaphor($F(1, 238) = 1.333, p = .249, \eta_{p}^{2} = 0.006$). Over health and finance, we found no significant main effect of metaphor type on perceived intelligence ($F(1, 238) = 0.447, p = .504, \eta_{p}^{2} = 0.002$) and no interaction effect of awareness level as well ($F(1, 238) = 0.376, p = .541, \eta_{p}^{2} = 0.002$). \autoref{table 6} presents means and standard deviations for each category of metaphors within their specific domains. \autoref{fig:5} indicates the distribution of the perceived intention to adopt scores. 

\begin{table}[h!]
\caption{The perceived intelligence (µ ± $\sigma$: mean and standard deviation) of human and non-human metaphors for the domains of health, finance, and overall.}
\label{table 6}
\begin{tabular}{llll}
\toprule
\multicolumn{4}{c}{\textbf{Perceived Intelligence}} \\ \midrule
 & \multicolumn{1}{c}{Health} & \multicolumn{1}{c}{Finance} & \multicolumn{1}{c}{Overall} \\
Human           & $\qty{4.42 \pm 0.57}{\percent}$       & $\qty{4.43 \pm 0.60}{\percent}$        & $\qty{4.43 \pm 0.54}{\percent}$      \\
Non-human       & $\qty{4.41 \pm 0.58}{\percent}$      & $\qty{4.40 \pm 0.59}{\percent}$        & $\qty{4.42 \pm 0.55}{\percent}$      \\ \bottomrule
\end{tabular}
\end{table}

In sum, our findings suggest that the human and non-human metaphors did not significantly impact perceived intelligence across all domains. There was also no significant effect of awareness level on perceptions of intelligence. 

\subsection{Qualitative Perceptions}

In addition to the questionnaires, we also gave the participants the option to provide their qualitative opinions of each version of `Z.' On average, 54.16\% of the participants provided some qualitative feedback on each metaphor across both awareness levels. The first author analyzed the qualitative data using an inductive thematic analysis approach. This involved generating initial codes from the data, which were then grouped into emerging themes to capture recurring patterns in participant responses.

For health, the participants considered the human metaphor (doctor) to be ``helpful,'' ``pleasant'' and "interactive" (mentioned by five participants each). On the other hand, the participants considered the non-human metaphor (encyclopedia) to be more ``informative,'' ``robotic,'' and ``knowledgeable'' (mentioned by eight, seven, and five participants, respectively). We found that participants had diverging opinions on how the human and non-human metaphors handled citing sources. Some participants found the human metaphor to be ``\textit{easier to listen to}'' (P235) because it did not mention sources, while other participants considered the non-human metaphor ``\textit{likelier to listen to}'' (P238) because it cited sources. One participant in implicit awareness level rightly identified the non-human metaphor as ``\textit{walking encyclopedia}'' (P221). However, for finance, participants perceived the human (financial advisor) and non-human metaphor (calculator) to be "friendly," "helpful," and "knowledgeable" (mentioned by five participants each). 

Interestingly, although we used the same voices (with the same pitch, tone, and volume) for all metaphors, some participants perceived the voices to be different based on the metaphor they were interacting with. For example, P68 described the voice of the calculator metaphor as ``\textit{this voice sounds more clinical and less friendly.}'' While P184 described the financial advisor metaphor as ``\textit{I felt genuine care from the voice of this version of `Z.'}'' The same finding persisted with the metaphors in the health domain as well. P30, in the feedback of the doctor metaphor, said, ``\textit{voice was more pleasant and human-like than the last one (encyclopedia).}'' Conversely, P66 and P103 described the encyclopedia metaphor voice as ``\textit{more robotic}'' and ``\textit{less personable}'' respectively. This finding highlights how the content and presentation of information can have an impact on how people perceive the voices of the VUIs. 

\section{Discussion}\label{discussion}

In this study, we found that the metaphorical nature of VUIs impacts user perceptions—however, there are caveats. Most importantly, the context of the interaction matters significantly. In the context of health, we found that the human metaphor was perceived to be likable and enjoyable than the non-human metaphor. Meanwhile, for finance, the metaphorical nature of the VUI did not influence perception. Moreover, regardless of the context, users found both human and non-human metaphors trustworthy and intelligent, and it did not influence their intentions to adopt the VUI. Surprisingly, we did not find much significant effect of awareness of the metaphor on user perceptions. However, participants perceived intending to adopt non-human metaphors slightly more if they were unaware of their non-human nature. 

 Although there is not much literature about designing non-human metaphors on user perceptions of VUIs, we do have some related studies to compare our findings within the space of chatbot design. Most prominently, Jung et al. \cite{Jung_Qiu_Bozzon_Gadiraju_2022} found that a `book' metaphor resulted in lower enjoyment and engagement of tasks compared to a `human' metaphor; however, the book metaphor also resulted in reduced cognitive load. Similarly, in our study, in the context of health, we found that the health encyclopedia metaphor was perceived to be less likable and enjoyable than the doctor metaphor. Some qualitative studies with VUIs could explain this finding in the context of health information seeking. Older adults, in co-design sessions, imagined their ideal VUI for imparting health information to be ``compassionate'' \cite{Desai_Lundy_Chin_2023, Desai_Hu_Lundy_Chin_2023}. Since receiving information about health could be sensitive, users prefer receiving this information in a more `humane' way. We designed the health encyclopedia VUI to be direct and informative, but we found that users did prefer dialogue elements that signaled humanness. For example, P219, in the implicit awareness condition, mentioned for the doctor metaphor, ``\textit{`I am here for you' is an oddly wholesome thing to hear at the end of the conversation and feels suitable (to this context).}'' Moreover, a doctor-patient relationship is more `personal' in nature \cite{Stokes_Dixon-Woods_McKinley_2004} and people prefer getting medical advice from humans than AI—even if AI outperforms doctors on various diagnosis metrics \cite{Longoni_Bonezzi_Morewedge_2019}. We see a carry-over of this preference in our study as well, wherein the measures related to likeability and enjoyment were more affected than trust, intelligence, or even perceived intention to adopt. 

Conversely, for finance, we did not find differences in how participants perceived human and non-human metaphors. This finding aligns with research in AI-based recommendation literature for finance-based decision-making. Users typically perceive advice from AI favorably in these scenarios. A recent survey found that about 31\% of users would accept recommendations from conversational AI, like ChatGPT, without verifying them.\footnote{\url{https://www.cnbc.com/2023/08/24/31percent-of-investors-are-ok-with-using-ai-as-their-financial-advisor.html}} Additionally, positive attitude towards AI correlates with intention to accept advice from a CUI. In our study, we found some participants had strong opinions on getting financial advice from the VUI. P140, in explicit condition, said about the calculator metaphor, ``\textit{I definitely would not be telling my voice interface information regarding my debts, income, or more. I feel providing that information to something not within your direct control is unwise. Additionally, the voice interface asking further questions regarding your financial situation is also unnerving.}'' They reiterated their opinion on the financial advisor metaphor as well. However, overall, the metaphorical nature of VUIs did not influence perceptions or perceived adoption. 

Desai \& Twidale \cite{Desai_Twidale_2023}, in their framework for metaphor contextualization, found a great imbalance between deployments of VUIs in human and non-human roles. Although users perceived VUIs to be human, non-human, or something in between \cite{Pradhan_Findlater_Lazar_2019, 10.1145/3322276.3322332}, manufacturers, designers, and researchers continuously conceptualize, design, and present VUIs in social roles. This gulf between how the users perceive VUIs and what is presented to them is not rooted in scientific reasoning and is more aspirational than practical \cite{Edlund_2019}. In their paper, aptly titled ``Once a Kind Friend is Now a Thing,'' Cho et al. \cite{10.1145/3322276.3322332} discuss how user expectations for VUIs change longitudinally. Commercial VUIs are advertised as ``assistants,'' so naturally, users expect human-like efficiency from them. However, with long-term use, these expectations decrease, and eventually, users adapt their mental models for VUIs—relegating them from being friends to things. However, in this paper, we challenge conventional wisdom in conversation design by asking—what if we design VUIs as ``\textit{things}''? Our findings highlight that human metaphors, although preferred in some contexts, are not a panacea. More nuance is needed when designing VUIs, and practitioners should consider if a non-human metaphor might be better for their use case. 

\subsection{Design Implications}

Our findings have several implications for design. We will distill these implications into three main design guidelines (DG) to help conversation designers identify the right metaphor for their design. 

\subsubsection{DG1: One Size `Does Not' Fit All}

Currently, VUIs are designed to occupy \textit{one} metaphor or persona. For example, Siri acts like an assistant, regardless of the context of the conversation. The tone or delivery style of commercial VUIs does not change based on user request. Responses to queries like ``Do I have Alzheimer's?'' will be answered in the same manner as ``Should I invest in crypto?'' We consider this to be absurd and problematic. Results from this study show that a human metaphor is more likeable and enjoyable for sensitive contexts like health than finance. So, conversation designers should consider the sensitivity of the context in which their VUI will be deployed in designing their personas. Although the inclination towards designing human personas is understandable (given how prevalent they are in design), the decision should be rooted in the context of eventual use. Moreover, research shows that CUIs designed using human metaphors set high expectations as users perceive them to be more competent, but with subsequent use, these highly competent metaphors are rejected for metaphors that project less competence \cite{Khadpe_Krishna_Fei-Fei_Hancock_Bernstein_2020}. Similarly, non-human metaphors might not project high competence \cite{Jung_Kim_So_Kim_Oh_2019} or seem more familiar; they will set more realistic expectations and are more aligned with users' mental models for VUIs \cite{Desai_Twidale_2022}.

\subsubsection{DG2: Don't Show, Don't Tell}

Studies involving metaphorical CUIs have traditionally involved chatbots \cite{Jung_Kim_So_Kim_Oh_2019, Khadpe_Krishna_Fei-Fei_Hancock_Bernstein_2020}. With chatbots, there are several ways in which one can signal the metaphorical nature of the interface, including the avatar (or an associated icon) of the CUI. However, with VUIs, there is no graphical medium to denote the metaphor without explicitly telling the user that they are interacting with a VUI modeled after a specific metaphor. Nonetheless, we found that the awareness of the metaphorical nature of the VUI did not have as much impact on user perceptions as we anticipated. Further, even in our qualitative data, we did not find much mention of how the metaphor influenced participants' opinions of the VUIs. We did observe some detrimental effects of awareness on the intention to adopt non-human metaphors. Therefore, we advise conversation designers to employ system personas as a strategic element in VUI design while discreetly managing these choices from end-users, particularly when integrating non-human metaphors into their designs.

\subsubsection{DG3: Be like a Chameleon}

In GUIs, a primary desktop metaphor is decomposed into secondary and auxiliary metaphors (e.g., files, folders, menus, bins, etc.). However, for VUIs, there is merely a primary `humanness' metaphor. Of course, a humanness metaphor is also somewhat nuanced because it gives rise to questions like `What kind of human?,' `doing what?' Yet it is difficult to base technologies on humans because humans are \textit{very} complicated. People adapt their behavior based on the context they are in or who they are interacting with. Currently, in our designs, we do not account for these complexities or contextual awareness. However, our results advocate for these complexities to be taken into account for design. We conceptualize a design space where the metaphor presented to the user changes based on the context of the conversation. If a user asks for health information, a VUI could use a human doctor metaphor, and if a user asks for financial advice, it can act like a calculator. The key idea is to adapt the metaphor to the conversational context. In this study, we used two human metaphors and two non-human metaphors and compared our results between them. However, the use of these specific metaphors is not as interesting as what the design implications are. We urge researchers to experiment with other metaphors in various contexts or even mix metaphors altogether. A playful approach to design is needed, and the rigid structures of system persona-based design cannot achieve it. 

\subsection{Limitations and Future work}
In the current study, participants formed their impressions of VUIs by examining hypothetical dialogues between a user and a metaphorical VUI. While this method is common practice in VUI research \cite{Seaborn_Urakami_2021}, it notably lacks the inclusion of direct interaction with the technology, offering insights that are observational rather than experiential. Further, perceptions of VUIs, along with the users' relationship with them, change over time \cite{10.1145/3322276.3322332}. Therefore, discerning how the outcomes of this study might translate to or differ in longitudinal research settings is important. Moreover, we restricted our study to native English speakers living in the U.S. This was done because the use of metaphors depends on cultural factors \cite{Moser_2000}, which could introduce additional confounding variables to this study. Going forward, we encourage researchers to study how cultural understanding of human and non-human metaphors influences perception. Additionally, longitudinal studies gauging perceptions of using metaphorical VUIs would extend our understanding of the applications of non-human metaphors in conversation design. 

The aim of this study is to introduce the possibility of designing VUIs using non-human metaphors to conversation designers. We present this study as the foundation from which the discourse can expand, acknowledging the numerous pathways for advancing this line of inquiry further:

\begin{itemize}
    \item \textbf{\textit{Organizing workshops}} will give us an opportunity to bring an interdisciplinary community together, encompassing conversation designers, cognitive psychologists, and developers to continue discussion regarding the importance of metaphors in VUI design. 
    \item \textbf{\textit{Longitudinal studies}} are necessary to assess if the observed impact of different metaphors persists over time and evaluate how it differs based on the context of the interaction (e.g., `lab' vs. `in-the-wild' studies). 
    \item \textbf{\textit{Working VUI prototypes}} will be developed based on different metaphors in collaboration with colleagues from industry. The aim of developing these prototypes will be to improve the ecological validity and obtain more realistic interaction insights that more closely reflect the real-life interactions between users and VUIs.  
    \item \textbf{\textit{Different types of voices}} will be synthesized and evaluated. Specifically, we will experiment with different genders and accents of voices to investigate the suitability of voices with different prosodic qualities (pitch, tempo, volume, etc.) on audiences that differ in terms of gender and cultural background.
    \item \textbf{\textit{Different metaphors}} that go beyond the realm of finance and health will be explored. We will experiment with domains such as fantasy characters (e.g., `Gandalf', `Jar Jar Binks') and focus on the metaphors that are relatable to specific audiences (e.g., `Lord of the Rings' fans, `Star Wars' community).
 \end{itemize}

We hope that the insights gathered via the above action points will help researchers and conversation designers to create VUIs that are more suitable and better tailored for specific interaction contexts, potentially improving user experience and increasing the likelihood of adoption.



\bibliographystyle{ACM-Reference-Format}
\bibliography{Manuscript}


\begin{thebibliography}{82}


\ifx \showCODEN    \undefined \def \showCODEN     #1{\unskip}     \fi
\ifx \showDOI      \undefined \def \showDOI       #1{#1}\fi
\ifx \showISBNx    \undefined \def \showISBNx     #1{\unskip}     \fi
\ifx \showISBNxiii \undefined \def \showISBNxiii  #1{\unskip}     \fi
\ifx \showISSN     \undefined \def \showISSN      #1{\unskip}     \fi
\ifx \showLCCN     \undefined \def \showLCCN      #1{\unskip}     \fi
\ifx \shownote     \undefined \def \shownote      #1{#1}          \fi
\ifx \showarticletitle \undefined \def \showarticletitle #1{#1}   \fi
\ifx \showURL      \undefined \def \showURL       {\relax}        \fi
\providecommand\bibfield[2]{#2}
\providecommand\bibinfo[2]{#2}
\providecommand\natexlab[1]{#1}
\providecommand\showeprint[2][]{arXiv:#2}

\bibitem[Alnefaie et~al\mbox{.}(2021)]%
        {alnefaie2021overview}
\bibfield{author}{\bibinfo{person}{Ahlam Alnefaie}, \bibinfo{person}{Sonika Singh}, \bibinfo{person}{Baki Kocaballi}, {and} \bibinfo{person}{Mukesh Prasad}.} \bibinfo{year}{2021}\natexlab{}.
\newblock \showarticletitle{An overview of conversational agent: applications, challenges and future directions}. In \bibinfo{booktitle}{\emph{17th International Conference on Web Information Systems and Technologies}}. SCITEPRESS-Science and Technology Publications.
\newblock


\bibitem[Ammari et~al\mbox{.}(2019)]%
        {ammari2019music}
\bibfield{author}{\bibinfo{person}{Tawfiq Ammari}, \bibinfo{person}{Jofish Kaye}, \bibinfo{person}{Janice~Y Tsai}, {and} \bibinfo{person}{Frank Bentley}.} \bibinfo{year}{2019}\natexlab{}.
\newblock \showarticletitle{Music, search, and IoT: How people (really) use voice assistants}.
\newblock \bibinfo{journal}{\emph{ACM Transactions on Computer-Human Interaction (TOCHI)}} \bibinfo{volume}{26}, \bibinfo{number}{3} (\bibinfo{year}{2019}), \bibinfo{pages}{1--28}.
\newblock


\bibitem[Araujo(2018)]%
        {araujo2018living}
\bibfield{author}{\bibinfo{person}{Theo Araujo}.} \bibinfo{year}{2018}\natexlab{}.
\newblock \showarticletitle{Living up to the chatbot hype: The influence of anthropomorphic design cues and communicative agency framing on conversational agent and company perceptions}.
\newblock \bibinfo{journal}{\emph{Computers in Human Behavior}}  \bibinfo{volume}{85} (\bibinfo{year}{2018}), \bibinfo{pages}{183--189}.
\newblock


\bibitem[Bansal et~al\mbox{.}(2010)]%
        {bansal2010impact}
\bibfield{author}{\bibinfo{person}{Gaurav Bansal}, \bibinfo{person}{David Gefen}, {et~al\mbox{.}}} \bibinfo{year}{2010}\natexlab{}.
\newblock \showarticletitle{The impact of personal dispositions on information sensitivity, privacy concern and trust in disclosing health information online}.
\newblock \bibinfo{journal}{\emph{Decision support systems}} \bibinfo{volume}{49}, \bibinfo{number}{2} (\bibinfo{year}{2010}), \bibinfo{pages}{138--150}.
\newblock


\bibitem[Bartneck(2023)]%
        {Bartneck_2023}
\bibfield{author}{\bibinfo{person}{Christoph Bartneck}.} \bibinfo{year}{2023}\natexlab{}.
\newblock \bibinfo{booktitle}{\emph{Godspeed Questionnaire Series: Translations and Usage}}.
\newblock \bibinfo{publisher}{Springer International Publishing}, \bibinfo{address}{Cham}, \bibinfo{pages}{1–35}.
\newblock
\showISBNx{978-3-030-89738-3}
\urldef\tempurl%
\url{https://doi.org/10.1007/978-3-030-89738-3_24-1}
\showDOI{\tempurl}


\bibitem[Bartneck et~al\mbox{.}(2009)]%
        {Bartneck_Kulić_Croft_Zoghbi_2009}
\bibfield{author}{\bibinfo{person}{Christoph Bartneck}, \bibinfo{person}{Dana Kulić}, \bibinfo{person}{Elizabeth Croft}, {and} \bibinfo{person}{Susana Zoghbi}.} \bibinfo{year}{2009}\natexlab{}.
\newblock \showarticletitle{Measurement Instruments for the Anthropomorphism, Animacy, Likeability, Perceived Intelligence, and Perceived Safety of Robots}.
\newblock \bibinfo{journal}{\emph{International Journal of Social Robotics}} \bibinfo{volume}{1}, \bibinfo{number}{1} (\bibinfo{date}{Jan.} \bibinfo{year}{2009}), \bibinfo{pages}{71–81}.
\newblock
\showISSN{1875-4805}
\urldef\tempurl%
\url{https://doi.org/10.1007/s12369-008-0001-3}
\showDOI{\tempurl}


\bibitem[Bickmore et~al\mbox{.}(2009)]%
        {bickmore2009taking}
\bibfield{author}{\bibinfo{person}{Timothy~W Bickmore}, \bibinfo{person}{Laura~M Pfeifer}, {and} \bibinfo{person}{Brian~W Jack}.} \bibinfo{year}{2009}\natexlab{}.
\newblock \showarticletitle{Taking the time to care: empowering low health literacy hospital patients with virtual nurse agents}. In \bibinfo{booktitle}{\emph{Proceedings of the SIGCHI conference on human factors in computing systems}}. \bibinfo{pages}{1265--1274}.
\newblock


\bibitem[Blackwell(2006)]%
        {Blackwell_2006}
\bibfield{author}{\bibinfo{person}{Alan~F. Blackwell}.} \bibinfo{year}{2006}\natexlab{}.
\newblock \showarticletitle{The reification of metaphor as a design tool}.
\newblock \bibinfo{journal}{\emph{ACM Transactions on Computer-Human Interaction}} \bibinfo{volume}{13}, \bibinfo{number}{4} (\bibinfo{date}{Dec.} \bibinfo{year}{2006}), \bibinfo{pages}{490–530}.
\newblock
\showISSN{1073-0516}
\urldef\tempurl%
\url{https://doi.org/10.1145/1188816.1188820}
\showDOI{\tempurl}


\bibitem[Braun et~al\mbox{.}(2019)]%
        {Braun_Mainz_Chadowitz_Pfleging_Alt_2019}
\bibfield{author}{\bibinfo{person}{Michael Braun}, \bibinfo{person}{Anja Mainz}, \bibinfo{person}{Ronee Chadowitz}, \bibinfo{person}{Bastian Pfleging}, {and} \bibinfo{person}{Florian Alt}.} \bibinfo{year}{2019}\natexlab{}.
\newblock \showarticletitle{At Your Service: Designing Voice Assistant Personalities to Improve Automotive User Interfaces}. In \bibinfo{booktitle}{\emph{Proceedings of the 2019 CHI Conference on Human Factors in Computing Systems}}. \bibinfo{publisher}{ACM}, \bibinfo{address}{Glasgow Scotland Uk}, \bibinfo{pages}{1–11}.
\newblock
\showISBNx{978-1-4503-5970-2}
\urldef\tempurl%
\url{https://doi.org/10.1145/3290605.3300270}
\showDOI{\tempurl}


\bibitem[Cameron and Maslen(2010)]%
        {Cameron_Maslen_2010}
\bibfield{author}{\bibinfo{person}{Lynne Cameron} {and} \bibinfo{person}{Robert Maslen}.} \bibinfo{year}{2010}\natexlab{}.
\newblock \bibinfo{booktitle}{\emph{Metaphor Analysis: Research Practice in applied Linguistics, Social Sciences and the Humanities}}.
\newblock \bibinfo{publisher}{Equinox}, \bibinfo{address}{London}.
\newblock
\showISBNx{978-1-84553-446-2}
\urldef\tempurl%
\url{http://www.equinoxpub.com/books/showbook.asp?bkid=363&keyword=9781845534479}
\showURL{%
\tempurl}


\bibitem[Candello et~al\mbox{.}(2017)]%
        {candello2017shaping}
\bibfield{author}{\bibinfo{person}{Heloisa Candello}, \bibinfo{person}{Claudio Pinhanez}, \bibinfo{person}{David Millen}, {and} \bibinfo{person}{Bruna~Daniele Andrade}.} \bibinfo{year}{2017}\natexlab{}.
\newblock \showarticletitle{Shaping the experience of a cognitive investment adviser}. In \bibinfo{booktitle}{\emph{Design, User Experience, and Usability: Understanding Users and Contexts: 6th International Conference, DUXU 2017, Held as Part of HCI International 2017, Vancouver, BC, Canada, July 9-14, 2017, Proceedings, Part III 6}}. Springer, \bibinfo{pages}{594--613}.
\newblock


\bibitem[Cheng et~al\mbox{.}(2018)]%
        {cheng2018development}
\bibfield{author}{\bibinfo{person}{Amy Cheng}, \bibinfo{person}{Vaishnavi Raghavaraju}, \bibinfo{person}{Jayanth Kanugo}, \bibinfo{person}{Yohanes~P Handrianto}, {and} \bibinfo{person}{Yi Shang}.} \bibinfo{year}{2018}\natexlab{}.
\newblock \showarticletitle{Development and evaluation of a healthy coping voice interface application using the Google home for elderly patients with type 2 diabetes}. In \bibinfo{booktitle}{\emph{2018 15th IEEE Annual Consumer Communications \& Networking Conference (CCNC)}}. IEEE, \bibinfo{pages}{1--5}.
\newblock


\bibitem[Chin et~al\mbox{.}(2024)]%
        {Chin2024Like}
\bibfield{author}{\bibinfo{person}{Jessie Chin}, \bibinfo{person}{Smit Desai}, \bibinfo{person}{Sheny Lin}, {and} \bibinfo{person}{Shannon Mej{\'i}a}.} \bibinfo{year}{2024}\natexlab{}.
\newblock \showarticletitle{Like my aunt Dorothy: Effects of conversational styles on perceptions, acceptance, and metaphorical descriptions of voice assistants during later adulthood}.
\newblock \bibinfo{journal}{\emph{Proceedings of the ACM on Human-Computer Interaction}} \bibinfo{volume}{8}, \bibinfo{number}{CSCW1} (\bibinfo{date}{Apr} \bibinfo{year}{2024}), \bibinfo{pages}{1--22}.
\newblock
\urldef\tempurl%
\url{https://doi.org/10.1145/3637365}
\showDOI{\tempurl}


\bibitem[Cho et~al\mbox{.}(2019)]%
        {10.1145/3322276.3322332}
\bibfield{author}{\bibinfo{person}{Minji Cho}, \bibinfo{person}{Sang-su Lee}, {and} \bibinfo{person}{Kun-Pyo Lee}.} \bibinfo{year}{2019}\natexlab{}.
\newblock \showarticletitle{Once a Kind Friend is Now a Thing: Understanding How Conversational Agents at Home are Forgotten}. In \bibinfo{booktitle}{\emph{Proceedings of the 2019 on Designing Interactive Systems Conference}} (San Diego, CA, USA) \emph{(\bibinfo{series}{DIS '19})}. \bibinfo{publisher}{Association for Computing Machinery}, \bibinfo{address}{New York, NY, USA}, \bibinfo{pages}{1557–1569}.
\newblock
\showISBNx{9781450358507}
\urldef\tempurl%
\url{https://doi.org/10.1145/3322276.3322332}
\showDOI{\tempurl}


\bibitem[Colburn and Shute(2008)]%
        {Colburn_Shute_2008}
\bibfield{author}{\bibinfo{person}{T.~R. Colburn} {and} \bibinfo{person}{G.~M. Shute}.} \bibinfo{year}{2008}\natexlab{}.
\newblock \showarticletitle{Metaphor in computer science}.
\newblock \bibinfo{journal}{\emph{Journal of Applied Logic}} \bibinfo{volume}{6}, \bibinfo{number}{4} (\bibinfo{date}{Dec.} \bibinfo{year}{2008}), \bibinfo{pages}{526–533}.
\newblock
\showISSN{1570-8683}
\urldef\tempurl%
\url{https://doi.org/10.1016/j.jal.2008.09.005}
\showDOI{\tempurl}


\bibitem[Cowan et~al\mbox{.}(2017)]%
        {10.1145/3098279.3098539}
\bibfield{author}{\bibinfo{person}{Benjamin~R. Cowan}, \bibinfo{person}{Nadia Pantidi}, \bibinfo{person}{David Coyle}, \bibinfo{person}{Kellie Morrissey}, \bibinfo{person}{Peter Clarke}, \bibinfo{person}{Sara Al-Shehri}, \bibinfo{person}{David Earley}, {and} \bibinfo{person}{Natasha Bandeira}.} \bibinfo{year}{2017}\natexlab{}.
\newblock \showarticletitle{"What can i help you with?": infrequent users' experiences of intelligent personal assistants}. In \bibinfo{booktitle}{\emph{Proceedings of the 19th International Conference on Human-Computer Interaction with Mobile Devices and Services}} (Vienna, Austria) \emph{(\bibinfo{series}{MobileHCI '17})}. \bibinfo{publisher}{Association for Computing Machinery}, \bibinfo{address}{New York, NY, USA}, Article \bibinfo{articleno}{43}, \bibinfo{numpages}{12}~pages.
\newblock
\showISBNx{9781450350754}
\urldef\tempurl%
\url{https://doi.org/10.1145/3098279.3098539}
\showDOI{\tempurl}


\bibitem[Cox and Ooi(2022)]%
        {cox2022does}
\bibfield{author}{\bibinfo{person}{Samuel~Rhys Cox} {and} \bibinfo{person}{Wei~Tsang Ooi}.} \bibinfo{year}{2022}\natexlab{}.
\newblock \showarticletitle{Does Chatbot Language Formality Affect Users’ Self-Disclosure?}. In \bibinfo{booktitle}{\emph{Proceedings of the 4th Conference on Conversational User Interfaces}}. \bibinfo{pages}{1--13}.
\newblock


\bibitem[Curry et~al\mbox{.}(2020)]%
        {curry2020conversational}
\bibfield{author}{\bibinfo{person}{Amanda~Cercas Curry}, \bibinfo{person}{Judy Robertson}, {and} \bibinfo{person}{Verena Rieser}.} \bibinfo{year}{2020}\natexlab{}.
\newblock \showarticletitle{Conversational assistants and gender stereotypes: Public perceptions and desiderata for voice personas}. In \bibinfo{booktitle}{\emph{Proceedings of the second workshop on gender bias in natural language processing}}. \bibinfo{pages}{72--78}.
\newblock


\bibitem[Day et~al\mbox{.}(2018)]%
        {day2018artificial}
\bibfield{author}{\bibinfo{person}{Min-Yuh Day}, \bibinfo{person}{Jian-Ting Lin}, {and} \bibinfo{person}{Yuan-Chih Chen}.} \bibinfo{year}{2018}\natexlab{}.
\newblock \showarticletitle{Artificial intelligence for conversational robo-advisor}. In \bibinfo{booktitle}{\emph{2018 IEEE/ACM International Conference on Advances in Social Networks Analysis and Mining (ASONAM)}}. IEEE, \bibinfo{pages}{1057--1064}.
\newblock


\bibitem[Desai and Chin(2023)]%
        {Desai_Chin_2023}
\bibfield{author}{\bibinfo{person}{Smit Desai} {and} \bibinfo{person}{Jessie Chin}.} \bibinfo{year}{2023}\natexlab{}.
\newblock \showarticletitle{OK Google, Let’s Learn: Using Voice User Interfaces for Informal Self-Regulated Learning of Health Topics among Younger and Older Adults}. In \bibinfo{booktitle}{\emph{Proceedings of the 2023 CHI Conference on Human Factors in Computing Systems}} \emph{(\bibinfo{series}{CHI ’23})}. \bibinfo{publisher}{Association for Computing Machinery}, \bibinfo{address}{New York, NY, USA}, \bibinfo{pages}{1–21}.
\newblock
\showISBNx{978-1-4503-9421-5}
\urldef\tempurl%
\url{https://doi.org/10.1145/3544548.3581507}
\showDOI{\tempurl}


\bibitem[Desai et~al\mbox{.}(2023a)]%
        {Desai_Hu_Lundy_Chin_2023}
\bibfield{author}{\bibinfo{person}{Smit Desai}, \bibinfo{person}{Xinhui Hu}, \bibinfo{person}{Morgan Lundy}, {and} \bibinfo{person}{Jessie Chin}.} \bibinfo{year}{2023}\natexlab{a}.
\newblock \showarticletitle{Using Experience-Based Participatory Approach to Design Interactive Voice User Interfaces for Delivering Physical Activity Programs with Older Adults}. In \bibinfo{booktitle}{\emph{Proceedings of the 11th International Conference on Human-Agent Interaction}} \emph{(\bibinfo{series}{HAI ’23})}. \bibinfo{publisher}{Association for Computing Machinery}, \bibinfo{address}{New York, NY, USA}, \bibinfo{pages}{180–190}.
\newblock
\showISBNx{9798400708244}
\urldef\tempurl%
\url{https://doi.org/10.1145/3623809.3623827}
\showDOI{\tempurl}


\bibitem[Desai et~al\mbox{.}(2023b)]%
        {Desai_Lundy_Chin_2023}
\bibfield{author}{\bibinfo{person}{Smit Desai}, \bibinfo{person}{Morgan Lundy}, {and} \bibinfo{person}{Jessie Chin}.} \bibinfo{year}{2023}\natexlab{b}.
\newblock \showarticletitle{“A Painless Way to Learn:” Designing an Interactive Storytelling Voice User Interface to Engage Older Adults in Informal Health Information Learning}. In \bibinfo{booktitle}{\emph{Proceedings of the 5th International Conference on Conversational User Interfaces}} \emph{(\bibinfo{series}{CUI ’23})}. \bibinfo{publisher}{Association for Computing Machinery}, \bibinfo{address}{New York, NY, USA}, \bibinfo{pages}{1–16}.
\newblock
\showISBNx{9798400700149}
\urldef\tempurl%
\url{https://doi.org/10.1145/3571884.3597141}
\showDOI{\tempurl}


\bibitem[Desai and Twidale(2022)]%
        {Desai_Twidale_2022}
\bibfield{author}{\bibinfo{person}{Smit Desai} {and} \bibinfo{person}{Michael Twidale}.} \bibinfo{year}{2022}\natexlab{}.
\newblock \showarticletitle{Is Alexa like a computer? A search engine? A friend? A silly child? Yes.}. In \bibinfo{booktitle}{\emph{Proceedings of the 4th Conference on Conversational User Interfaces}} \emph{(\bibinfo{series}{CUI ’22})}. \bibinfo{publisher}{Association for Computing Machinery}, \bibinfo{address}{New York, NY, USA}, \bibinfo{pages}{1–4}.
\newblock
\showISBNx{978-1-4503-9739-1}
\urldef\tempurl%
\url{https://doi.org/10.1145/3543829.3544535}
\showDOI{\tempurl}


\bibitem[Desai and Twidale(2023)]%
        {Desai_Twidale_2023}
\bibfield{author}{\bibinfo{person}{Smit Desai} {and} \bibinfo{person}{Michael Twidale}.} \bibinfo{year}{2023}\natexlab{}.
\newblock \showarticletitle{Metaphors in Voice User Interfaces: A Slippery Fish}.
\newblock \bibinfo{journal}{\emph{ACM Transactions on Computer-Human Interaction}} \bibinfo{volume}{30}, \bibinfo{number}{6} (\bibinfo{date}{Dec.} \bibinfo{year}{2023}), \bibinfo{pages}{1–37}.
\newblock
\showISSN{1073-0516, 1557-7325}
\urldef\tempurl%
\url{https://doi.org/10.1145/3609326}
\showDOI{\tempurl}


\bibitem[DeVito et~al\mbox{.}(2018)]%
        {DeVito_Birnholtz_Hancock_French_Liu_2018}
\bibfield{author}{\bibinfo{person}{Michael~A. DeVito}, \bibinfo{person}{Jeremy Birnholtz}, \bibinfo{person}{Jeffery~T. Hancock}, \bibinfo{person}{Megan French}, {and} \bibinfo{person}{Sunny Liu}.} \bibinfo{year}{2018}\natexlab{}.
\newblock \showarticletitle{How People Form Folk Theories of Social Media Feeds and What it Means for How We Study Self-Presentation}. In \bibinfo{booktitle}{\emph{Proceedings of the 2018 CHI Conference on Human Factors in Computing Systems}} \emph{(\bibinfo{series}{CHI ’18})}. \bibinfo{publisher}{Association for Computing Machinery}, \bibinfo{address}{New York, NY, USA}, \bibinfo{pages}{1–12}.
\newblock
\showISBNx{978-1-4503-5620-6}
\urldef\tempurl%
\url{https://doi.org/10.1145/3173574.3173694}
\showDOI{\tempurl}


\bibitem[Dubiel et~al\mbox{.}(2020)]%
        {dubiel2020persuasive}
\bibfield{author}{\bibinfo{person}{Mateusz Dubiel}, \bibinfo{person}{Martin Halvey}, \bibinfo{person}{Pilar~Oplustil Gallegos}, {and} \bibinfo{person}{Simon King}.} \bibinfo{year}{2020}\natexlab{}.
\newblock \showarticletitle{Persuasive synthetic speech: Voice perception and user behaviour}. In \bibinfo{booktitle}{\emph{Proceedings of the 2nd Conference on Conversational User Interfaces}}. \bibinfo{pages}{1--9}.
\newblock


\bibitem[Dubiel et~al\mbox{.}(2024)]%
        {dubiel2024impact}
\bibfield{author}{\bibinfo{person}{Mateusz Dubiel}, \bibinfo{person}{Anastasia Sergeeva}, {and} \bibinfo{person}{Luis~A. Leiva}.} \bibinfo{year}{2024}\natexlab{}.
\newblock \bibinfo{title}{Impact of Voice Fidelity on Decision Making: A Potential Dark Pattern?}
\newblock
\newblock
\showeprint[arxiv]{2402.07010}~[cs.HC]


\bibitem[Edlund(2019)]%
        {Edlund_2019}
\bibfield{author}{\bibinfo{person}{Jens Edlund}.} \bibinfo{year}{2019}\natexlab{}.
\newblock \showarticletitle{Shoehorning in the name of science}. In \bibinfo{booktitle}{\emph{Proceedings of the 1st International Conference on Conversational User Interfaces}} \emph{(\bibinfo{series}{CUI ’19})}. \bibinfo{publisher}{Association for Computing Machinery}, \bibinfo{address}{New York, NY, USA}, \bibinfo{pages}{1–3}.
\newblock
\showISBNx{978-1-4503-7187-2}
\urldef\tempurl%
\url{https://doi.org/10.1145/3342775.3342794}
\showDOI{\tempurl}


\bibitem[Fadhil et~al\mbox{.}(2019)]%
        {fadhil2019assistive}
\bibfield{author}{\bibinfo{person}{Ahmed Fadhil}, \bibinfo{person}{Yunlong Wang}, {and} \bibinfo{person}{Harald Reiterer}.} \bibinfo{year}{2019}\natexlab{}.
\newblock \showarticletitle{Assistive conversational agent for health coaching: a validation study}.
\newblock \bibinfo{journal}{\emph{Methods of information in medicine}} \bibinfo{volume}{58}, \bibinfo{number}{01} (\bibinfo{year}{2019}), \bibinfo{pages}{009--023}.
\newblock


\bibitem[Halperin et~al\mbox{.}(2023)]%
        {halperin2023probing}
\bibfield{author}{\bibinfo{person}{Brett~A Halperin}, \bibinfo{person}{Gary Hsieh}, \bibinfo{person}{Erin McElroy}, \bibinfo{person}{James Pierce}, {and} \bibinfo{person}{Daniela~K Rosner}.} \bibinfo{year}{2023}\natexlab{}.
\newblock \showarticletitle{Probing a Community-Based Conversational Storytelling Agent to Document Digital Stories of Housing Insecurity}. In \bibinfo{booktitle}{\emph{Proceedings of the 2023 CHI Conference on Human Factors in Computing Systems}}. \bibinfo{pages}{1--18}.
\newblock


\bibitem[Heckel(1984)]%
        {Heckel_1984}
\bibfield{author}{\bibinfo{person}{Paul Heckel}.} \bibinfo{year}{1984}\natexlab{}.
\newblock \bibinfo{booktitle}{\emph{The elements of friendly software design}}.
\newblock \bibinfo{publisher}{New York: Warner Books}.
\newblock
\showISBNx{978-0-446-38040-9}
\urldef\tempurl%
\url{http://archive.org/details/elementsoffriend00heck}
\showURL{%
\tempurl}


\bibitem[Hildebrand and Bergner(2021)]%
        {hildebrand2021conversational}
\bibfield{author}{\bibinfo{person}{Christian Hildebrand} {and} \bibinfo{person}{Anouk Bergner}.} \bibinfo{year}{2021}\natexlab{}.
\newblock \showarticletitle{Conversational robo advisors as surrogates of trust: onboarding experience, firm perception, and consumer financial decision making}.
\newblock \bibinfo{journal}{\emph{Journal of the Academy of Marketing Science}}  \bibinfo{volume}{49} (\bibinfo{year}{2021}), \bibinfo{pages}{659--676}.
\newblock


\bibitem[Hodge et~al\mbox{.}(2021)]%
        {hodge2021effect}
\bibfield{author}{\bibinfo{person}{Frank~D Hodge}, \bibinfo{person}{Kim~I Mendoza}, {and} \bibinfo{person}{Roshan~K Sinha}.} \bibinfo{year}{2021}\natexlab{}.
\newblock \showarticletitle{The effect of humanizing robo-advisors on investor judgments}.
\newblock \bibinfo{journal}{\emph{Contemporary Accounting Research}} \bibinfo{volume}{38}, \bibinfo{number}{1} (\bibinfo{year}{2021}), \bibinfo{pages}{770--792}.
\newblock


\bibitem[Indurkhya(2013)]%
        {Indurkhya_2013}
\bibfield{author}{\bibinfo{person}{B. Indurkhya}.} \bibinfo{year}{2013}\natexlab{}.
\newblock \bibinfo{booktitle}{\emph{Metaphor and Cognition: An Interactionist Approach}}.
\newblock \bibinfo{publisher}{Springer Science \& Business Media}.
\newblock
\showISBNx{978-94-017-2252-0}
\newblock
\shownote{Google-Books-ID: foTrCAAAQBAJ}.


\bibitem[Jackendoff and Aaron(1991)]%
        {Jackendoff_Aaron_1991}
\bibfield{author}{\bibinfo{person}{Ray Jackendoff} {and} \bibinfo{person}{David Aaron}.} \bibinfo{year}{1991}\natexlab{}.
\newblock \showarticletitle{Review of More Than Cool Reason: A Field Guide to Poetic Metaphor}.
\newblock \bibinfo{journal}{\emph{Language}} \bibinfo{volume}{67}, \bibinfo{number}{2} (\bibinfo{year}{1991}), \bibinfo{pages}{320–338}.
\newblock
\showISSN{0097-8507}
\urldef\tempurl%
\url{https://doi.org/10.2307/415109}
\showDOI{\tempurl}


\bibitem[Jian et~al\mbox{.}(2000)]%
        {Jian_Bisantz_Drury_2000}
\bibfield{author}{\bibinfo{person}{Jiun-Yin Jian}, \bibinfo{person}{Ann~M. Bisantz}, {and} \bibinfo{person}{Colin~G. Drury}.} \bibinfo{year}{2000}\natexlab{}.
\newblock \showarticletitle{Foundations for an empirically determined scale of trust in automated systems}.
\newblock \bibinfo{journal}{\emph{International Journal of Cognitive Ergonomics}} \bibinfo{volume}{4}, \bibinfo{number}{1} (\bibinfo{year}{2000}), \bibinfo{pages}{53–71}.
\newblock
\showISSN{1532-7566}
\urldef\tempurl%
\url{https://doi.org/10.1207/S15327566IJCE0401_04}
\showDOI{\tempurl}


\bibitem[Jung et~al\mbox{.}(2019)]%
        {Jung_Kim_So_Kim_Oh_2019}
\bibfield{author}{\bibinfo{person}{Hyunhoon Jung}, \bibinfo{person}{Hee~Jae Kim}, \bibinfo{person}{Seongeun So}, \bibinfo{person}{Jinjoong Kim}, {and} \bibinfo{person}{Changhoon Oh}.} \bibinfo{year}{2019}\natexlab{}.
\newblock \showarticletitle{TurtleTalk: An Educational Programming Game for Children with Voice User Interface}. In \bibinfo{booktitle}{\emph{Extended Abstracts of the 2019 CHI Conference on Human Factors in Computing Systems}} \emph{(\bibinfo{series}{CHI EA ’19})}. \bibinfo{publisher}{Association for Computing Machinery}, \bibinfo{address}{New York, NY, USA}, \bibinfo{pages}{1–6}.
\newblock
\showISBNx{978-1-4503-5971-9}
\urldef\tempurl%
\url{https://doi.org/10.1145/3290607.3312773}
\showDOI{\tempurl}


\bibitem[Jung et~al\mbox{.}(2022)]%
        {Jung_Qiu_Bozzon_Gadiraju_2022}
\bibfield{author}{\bibinfo{person}{Ji-Youn Jung}, \bibinfo{person}{Sihang Qiu}, \bibinfo{person}{Alessandro Bozzon}, {and} \bibinfo{person}{Ujwal Gadiraju}.} \bibinfo{year}{2022}\natexlab{}.
\newblock \showarticletitle{Great Chain of Agents: The Role of Metaphorical Representation of Agents in Conversational Crowdsourcing}. In \bibinfo{booktitle}{\emph{CHI Conference on Human Factors in Computing Systems}}. \bibinfo{publisher}{ACM}, \bibinfo{address}{New Orleans LA USA}, \bibinfo{pages}{1–22}.
\newblock
\showISBNx{978-1-4503-9157-3}
\urldef\tempurl%
\url{https://doi.org/10.1145/3491102.3517653}
\showDOI{\tempurl}


\bibitem[Khadpe et~al\mbox{.}(2020)]%
        {Khadpe_Krishna_Fei-Fei_Hancock_Bernstein_2020}
\bibfield{author}{\bibinfo{person}{Pranav Khadpe}, \bibinfo{person}{Ranjay Krishna}, \bibinfo{person}{Li Fei-Fei}, \bibinfo{person}{Jeffrey~T. Hancock}, {and} \bibinfo{person}{Michael~S. Bernstein}.} \bibinfo{year}{2020}\natexlab{}.
\newblock \showarticletitle{Conceptual Metaphors Impact Perceptions of Human-AI Collaboration}.
\newblock \bibinfo{journal}{\emph{Proceedings of the ACM on Human-Computer Interaction}} \bibinfo{volume}{4}, \bibinfo{number}{CSCW2} (\bibinfo{date}{Oct.} \bibinfo{year}{2020}), \bibinfo{pages}{1–26}.
\newblock
\showISSN{2573-0142}
\urldef\tempurl%
\url{https://doi.org/10.1145/3415234}
\showDOI{\tempurl}


\bibitem[Kim and Choudhury(2021)]%
        {Kim_Choudhury_2021}
\bibfield{author}{\bibinfo{person}{Sunyoung Kim} {and} \bibinfo{person}{Abhishek Choudhury}.} \bibinfo{year}{2021}\natexlab{}.
\newblock \showarticletitle{Exploring older adults’ perception and use of smart speaker-based voice assistants: A longitudinal study}.
\newblock \bibinfo{journal}{\emph{Computers in Human Behavior}}  \bibinfo{volume}{124} (\bibinfo{date}{Nov.} \bibinfo{year}{2021}), \bibinfo{pages}{106914}.
\newblock
\showISSN{0747-5632}
\urldef\tempurl%
\url{https://doi.org/10.1016/j.chb.2021.106914}
\showDOI{\tempurl}


\bibitem[Kim et~al\mbox{.}(2023)]%
        {Kim_Molina_Rheu_Zhan_Peng_2023}
\bibfield{author}{\bibinfo{person}{Taenyun Kim}, \bibinfo{person}{Maria~D. Molina}, \bibinfo{person}{Minjin~(MJ) Rheu}, \bibinfo{person}{Emily~S. Zhan}, {and} \bibinfo{person}{Wei Peng}.} \bibinfo{year}{2023}\natexlab{}.
\newblock \showarticletitle{One AI Does Not Fit All: A Cluster Analysis of the Laypeople’s Perception of AI Roles}. In \bibinfo{booktitle}{\emph{Proceedings of the 2023 CHI Conference on Human Factors in Computing Systems}} \emph{(\bibinfo{series}{CHI ’23})}. \bibinfo{publisher}{Association for Computing Machinery}, \bibinfo{address}{New York, NY, USA}, \bibinfo{pages}{1–20}.
\newblock
\showISBNx{978-1-4503-9421-5}
\urldef\tempurl%
\url{https://doi.org/10.1145/3544548.3581340}
\showDOI{\tempurl}


\bibitem[Kocielnik et~al\mbox{.}(2021)]%
        {kocielnik2021can}
\bibfield{author}{\bibinfo{person}{Rafal Kocielnik}, \bibinfo{person}{Raina Langevin}, \bibinfo{person}{James~S George}, \bibinfo{person}{Shota Akenaga}, \bibinfo{person}{Amelia Wang}, \bibinfo{person}{Darwin~P Jones}, \bibinfo{person}{Alexander Argyle}, \bibinfo{person}{Callan Fockele}, \bibinfo{person}{Layla Anderson}, \bibinfo{person}{Dennis~T Hsieh}, {et~al\mbox{.}}} \bibinfo{year}{2021}\natexlab{}.
\newblock \showarticletitle{Can I Talk to You about Your Social Needs? Understanding Preference for Conversational User Interface in Health}. In \bibinfo{booktitle}{\emph{Proceedings of the 3rd Conference on Conversational User Interfaces}}. \bibinfo{pages}{1--10}.
\newblock


\bibitem[Kuzminykh et~al\mbox{.}(2020)]%
        {Kuzminykh_Sun_Govindaraju_Avery_Lank_2020}
\bibfield{author}{\bibinfo{person}{Anastasia Kuzminykh}, \bibinfo{person}{Jenny Sun}, \bibinfo{person}{Nivetha Govindaraju}, \bibinfo{person}{Jeff Avery}, {and} \bibinfo{person}{Edward Lank}.} \bibinfo{year}{2020}\natexlab{}.
\newblock \showarticletitle{Genie in the Bottle: Anthropomorphized Perceptions of Conversational Agents}. In \bibinfo{booktitle}{\emph{Proceedings of the 2020 CHI Conference on Human Factors in Computing Systems}} \emph{(\bibinfo{series}{CHI ’20})}. \bibinfo{publisher}{Association for Computing Machinery}, \bibinfo{address}{New York, NY, USA}, \bibinfo{pages}{1–13}.
\newblock
\showISBNx{978-1-4503-6708-0}
\urldef\tempurl%
\url{https://doi.org/10.1145/3313831.3376665}
\showDOI{\tempurl}


\bibitem[Lakoff and Johnson(1980)]%
        {Lakoff_Johnson_1980}
\bibfield{author}{\bibinfo{person}{George Lakoff} {and} \bibinfo{person}{Mark Johnson}.} \bibinfo{year}{1980}\natexlab{}.
\newblock \bibinfo{booktitle}{\emph{Metaphors We Live By}}.
\newblock \bibinfo{publisher}{University of Chicago Press}.
\newblock
\showISBNx{978-0-226-47099-3}
\newblock
\shownote{Google-Books-ID: r6nOYYtxzUoC}.


\bibitem[Lee et~al\mbox{.}(2021)]%
        {Lee_Frank_IJsselsteijn_2021}
\bibfield{author}{\bibinfo{person}{Minha Lee}, \bibinfo{person}{Lily Frank}, {and} \bibinfo{person}{Wijnand IJsselsteijn}.} \bibinfo{year}{2021}\natexlab{}.
\newblock \showarticletitle{Brokerbot: A Cryptocurrency Chatbot in the Social-technical Gap of Trust}.
\newblock \bibinfo{journal}{\emph{Computer Supported Cooperative Work (CSCW)}} \bibinfo{volume}{30}, \bibinfo{number}{1} (\bibinfo{date}{Feb.} \bibinfo{year}{2021}), \bibinfo{pages}{79–117}.
\newblock
\showISSN{1573-7551}
\urldef\tempurl%
\url{https://doi.org/10.1007/s10606-021-09392-6}
\showDOI{\tempurl}


\bibitem[Liao and Vaughan(2023)]%
        {liao2023ai}
\bibfield{author}{\bibinfo{person}{Q.~Vera Liao} {and} \bibinfo{person}{Jennifer~Wortman Vaughan}.} \bibinfo{year}{2023}\natexlab{}.
\newblock \bibinfo{title}{AI Transparency in the Age of LLMs: A Human-Centered Research Roadmap}.
\newblock
\newblock
\showeprint[arxiv]{2306.01941}~[cs.HC]


\bibitem[Longoni et~al\mbox{.}(2019)]%
        {Longoni_Bonezzi_Morewedge_2019}
\bibfield{author}{\bibinfo{person}{Chiara Longoni}, \bibinfo{person}{Andrea Bonezzi}, {and} \bibinfo{person}{Carey~K Morewedge}.} \bibinfo{year}{2019}\natexlab{}.
\newblock \showarticletitle{Resistance to Medical Artificial Intelligence}.
\newblock \bibinfo{journal}{\emph{Journal of Consumer Research}} \bibinfo{volume}{46}, \bibinfo{number}{4} (\bibinfo{date}{Dec.} \bibinfo{year}{2019}), \bibinfo{pages}{629–650}.
\newblock
\showISSN{0093-5301}
\urldef\tempurl%
\url{https://doi.org/10.1093/jcr/ucz013}
\showDOI{\tempurl}


\bibitem[Luger and Sellen(2016)]%
        {Luger_Sellen_2016}
\bibfield{author}{\bibinfo{person}{Ewa Luger} {and} \bibinfo{person}{Abigail Sellen}.} \bibinfo{year}{2016}\natexlab{}.
\newblock \showarticletitle{“Like Having a Really Bad PA”: The Gulf between User Expectation and Experience of Conversational Agents}. In \bibinfo{booktitle}{\emph{Proceedings of the 2016 CHI Conference on Human Factors in Computing Systems}} \emph{(\bibinfo{series}{CHI ’16})}. \bibinfo{publisher}{Association for Computing Machinery}, \bibinfo{address}{New York, NY, USA}, \bibinfo{pages}{5286–5297}.
\newblock
\showISBNx{978-1-4503-3362-7}
\urldef\tempurl%
\url{https://doi.org/10.1145/2858036.2858288}
\showDOI{\tempurl}


\bibitem[Lupetti et~al\mbox{.}(2023)]%
        {lupetti2023trustworthy}
\bibfield{author}{\bibinfo{person}{Maria~Luce Lupetti}, \bibinfo{person}{Emma Hagens}, \bibinfo{person}{Willem Van Der~Maden}, \bibinfo{person}{R{\'e}gine Steegers-Theunissen}, {and} \bibinfo{person}{Melek Rousian}.} \bibinfo{year}{2023}\natexlab{}.
\newblock \showarticletitle{Trustworthy Embodied Conversational Agents for Healthcare: A Design Exploration of Embodied Conversational Agents for the periconception period at Erasmus MC}. In \bibinfo{booktitle}{\emph{Proceedings of the 5th International Conference on Conversational User Interfaces}}. \bibinfo{pages}{1--14}.
\newblock


\bibitem[Luxton(2020)]%
        {luxton2020ethical}
\bibfield{author}{\bibinfo{person}{David~D Luxton}.} \bibinfo{year}{2020}\natexlab{}.
\newblock \showarticletitle{Ethical implications of conversational agents in global public health}.
\newblock \bibinfo{journal}{\emph{Bulletin of the World Health Organization}} \bibinfo{volume}{98}, \bibinfo{number}{4} (\bibinfo{year}{2020}), \bibinfo{pages}{285}.
\newblock


\bibitem[McCrae and Costa(1986)]%
        {McCrae_Costa_1986}
\bibfield{author}{\bibinfo{person}{Robert~R. McCrae} {and} \bibinfo{person}{Paul~T. Costa}.} \bibinfo{year}{1986}\natexlab{}.
\newblock \showarticletitle{Personality, coping, and coping effectiveness in an adult sample}.
\newblock \bibinfo{journal}{\emph{Journal of Personality}} \bibinfo{volume}{54}, \bibinfo{number}{2} (\bibinfo{year}{1986}), \bibinfo{pages}{385–405}.
\newblock
\showISSN{1467-6494}
\urldef\tempurl%
\url{https://doi.org/10.1111/j.1467-6494.1986.tb00401.x}
\showDOI{\tempurl}


\bibitem[McMillan and Jaber(2021)]%
        {McMillan_Jaber_2021}
\bibfield{author}{\bibinfo{person}{Donald McMillan} {and} \bibinfo{person}{Razan Jaber}.} \bibinfo{year}{2021}\natexlab{}.
\newblock \showarticletitle{Leaving the Butler Behind: The Future of Role Reproduction in CUI}. In \bibinfo{booktitle}{\emph{CUI 2021 - 3rd Conference on Conversational User Interfaces}} \emph{(\bibinfo{series}{CUI ’21})}. \bibinfo{publisher}{Association for Computing Machinery}, \bibinfo{address}{New York, NY, USA}, \bibinfo{pages}{1–4}.
\newblock
\showISBNx{978-1-4503-8998-3}
\urldef\tempurl%
\url{https://doi.org/10.1145/3469595.3469606}
\showDOI{\tempurl}


\bibitem[Miner et~al\mbox{.}(2016)]%
        {miner2016smartphone}
\bibfield{author}{\bibinfo{person}{Adam~S Miner}, \bibinfo{person}{Arnold Milstein}, \bibinfo{person}{Stephen Schueller}, \bibinfo{person}{Roshini Hegde}, \bibinfo{person}{Christina Mangurian}, {and} \bibinfo{person}{Eleni Linos}.} \bibinfo{year}{2016}\natexlab{}.
\newblock \showarticletitle{Smartphone-based conversational agents and responses to questions about mental health, interpersonal violence, and physical health}.
\newblock \bibinfo{journal}{\emph{JAMA internal medicine}} \bibinfo{volume}{176}, \bibinfo{number}{5} (\bibinfo{year}{2016}), \bibinfo{pages}{619--625}.
\newblock


\bibitem[Morana et~al\mbox{.}(2020)]%
        {morana2020effect}
\bibfield{author}{\bibinfo{person}{Stefan Morana}, \bibinfo{person}{Ulrich Gnewuch}, \bibinfo{person}{Dominik Jung}, {and} \bibinfo{person}{Carsten Granig}.} \bibinfo{year}{2020}\natexlab{}.
\newblock \showarticletitle{The Effect of Anthropomorphism on Investment Decision-Making with Robo-Advisor Chatbots.}. In \bibinfo{booktitle}{\emph{ECIS}}.
\newblock


\bibitem[Moser(2000)]%
        {Moser_2000}
\bibfield{author}{\bibinfo{person}{Karin~S. Moser}.} \bibinfo{year}{2000}\natexlab{}.
\newblock \showarticletitle{Metaphor Analysis in Psychology—Method, Theory, and Fields of Application}.
\newblock \bibinfo{journal}{\emph{Forum Qualitative Sozialforschung / Forum: Qualitative Social Research}} \bibinfo{volume}{1}, \bibinfo{number}{22} (\bibinfo{date}{June} \bibinfo{year}{2000}).
\newblock
\showISSN{1438-5627}
\urldef\tempurl%
\url{https://doi.org/10.17169/fqs-1.2.1090}
\showDOI{\tempurl}


\bibitem[Motalebi et~al\mbox{.}(2019)]%
        {Motalebi_Cho_Sundar_Abdullah_2019}
\bibfield{author}{\bibinfo{person}{Nasim Motalebi}, \bibinfo{person}{Eugene Cho}, \bibinfo{person}{S.~Shyam Sundar}, {and} \bibinfo{person}{Saeed Abdullah}.} \bibinfo{year}{2019}\natexlab{}.
\newblock \showarticletitle{Can Alexa be your Therapist? How Back-Channeling Transforms Smart-Speakers to be Active Listeners}. In \bibinfo{booktitle}{\emph{Companion Publication of the 2019 Conference on Computer Supported Cooperative Work and Social Computing}} \emph{(\bibinfo{series}{CSCW ’19 Companion})}. \bibinfo{publisher}{Association for Computing Machinery}, \bibinfo{address}{New York, NY, USA}, \bibinfo{pages}{309–313}.
\newblock
\showISBNx{978-1-4503-6692-2}
\urldef\tempurl%
\url{https://doi.org/10.1145/3311957.3359502}
\showDOI{\tempurl}


\bibitem[Moussawi et~al\mbox{.}(2021)]%
        {Moussawi_Koufaris_Benbunan-Fich_2021}
\bibfield{author}{\bibinfo{person}{Sara Moussawi}, \bibinfo{person}{Marios Koufaris}, {and} \bibinfo{person}{Raquel Benbunan-Fich}.} \bibinfo{year}{2021}\natexlab{}.
\newblock \showarticletitle{How perceptions of intelligence and anthropomorphism affect adoption of personal intelligent agents}.
\newblock \bibinfo{journal}{\emph{Electronic Markets}} \bibinfo{volume}{31}, \bibinfo{number}{2} (\bibinfo{date}{June} \bibinfo{year}{2021}), \bibinfo{pages}{343–364}.
\newblock
\showISSN{1422-8890}
\urldef\tempurl%
\url{https://doi.org/10.1007/s12525-020-00411-w}
\showDOI{\tempurl}


\bibitem[Nass and Moon(2000)]%
        {Nass_Moon_2000}
\bibfield{author}{\bibinfo{person}{Clifford Nass} {and} \bibinfo{person}{Youngme Moon}.} \bibinfo{year}{2000}\natexlab{}.
\newblock \showarticletitle{Machines and Mindlessness: Social Responses to Computers}.
\newblock \bibinfo{journal}{\emph{Journal of Social Issues}} \bibinfo{volume}{56}, \bibinfo{number}{1} (\bibinfo{year}{2000}), \bibinfo{pages}{81–103}.
\newblock
\showISSN{1540-4560}
\urldef\tempurl%
\url{https://doi.org/10.1111/0022-4537.00153}
\showDOI{\tempurl}


\bibitem[Newman and Sproull(1979)]%
        {Newman_Sproull_1979}
\bibfield{author}{\bibinfo{person}{William~M. Newman} {and} \bibinfo{person}{Robert~F. Sproull}.} \bibinfo{year}{1979}\natexlab{}.
\newblock \bibinfo{booktitle}{\emph{Principles of Interactive Computer Graphics}}.
\newblock \bibinfo{publisher}{McGraw-Hill Education}, \bibinfo{address}{Auckland}.
\newblock
\showISBNx{978-0-07-066455-5}


\bibitem[Norman(2013)]%
        {Norman_2013}
\bibfield{author}{\bibinfo{person}{Don Norman}.} \bibinfo{year}{2013}\natexlab{}.
\newblock \bibinfo{booktitle}{\emph{The Design Of Everyday Things} (\bibinfo{edition}{revised edition} ed.)}.
\newblock \bibinfo{publisher}{Basic Books}, \bibinfo{address}{New York, New York}.
\newblock
\showISBNx{978-0-465-05065-9}


\bibitem[Porcheron et~al\mbox{.}(2018)]%
        {Porcheron_Fischer_Reeves_Sharples_2018}
\bibfield{author}{\bibinfo{person}{Martin Porcheron}, \bibinfo{person}{Joel~E. Fischer}, \bibinfo{person}{Stuart Reeves}, {and} \bibinfo{person}{Sarah Sharples}.} \bibinfo{year}{2018}\natexlab{}.
\newblock \showarticletitle{Voice Interfaces in Everyday Life}. In \bibinfo{booktitle}{\emph{Proceedings of the 2018 CHI Conference on Human Factors in Computing Systems}} \emph{(\bibinfo{series}{CHI ’18})}. \bibinfo{publisher}{Association for Computing Machinery}, \bibinfo{address}{New York, NY, USA}, \bibinfo{pages}{1–12}.
\newblock
\showISBNx{978-1-4503-5620-6}
\urldef\tempurl%
\url{https://doi.org/10.1145/3173574.3174214}
\showDOI{\tempurl}


\bibitem[Pradhan et~al\mbox{.}(2019)]%
        {Pradhan_Findlater_Lazar_2019}
\bibfield{author}{\bibinfo{person}{Alisha Pradhan}, \bibinfo{person}{Leah Findlater}, {and} \bibinfo{person}{Amanda Lazar}.} \bibinfo{year}{2019}\natexlab{}.
\newblock \showarticletitle{“Phantom Friend” or “Just a Box with Information”: Personification and Ontological Categorization of Smart Speaker-based Voice Assistants by Older Adults}.
\newblock \bibinfo{journal}{\emph{Proceedings of the ACM on Human-Computer Interaction}} \bibinfo{volume}{3}, \bibinfo{number}{CSCW} (\bibinfo{date}{Nov.} \bibinfo{year}{2019}), \bibinfo{pages}{1–21}.
\newblock
\showISSN{2573-0142}
\urldef\tempurl%
\url{https://doi.org/10.1145/3359316}
\showDOI{\tempurl}


\bibitem[Pradhan and Lazar(2021)]%
        {Pradhan_Lazar_2021}
\bibfield{author}{\bibinfo{person}{Alisha Pradhan} {and} \bibinfo{person}{Amanda Lazar}.} \bibinfo{year}{2021}\natexlab{}.
\newblock \showarticletitle{Hey Google, Do You Have a Personality? Designing Personality and Personas for Conversational Agents}. In \bibinfo{booktitle}{\emph{Proceedings of the 3rd Conference on Conversational User Interfaces}} \emph{(\bibinfo{series}{CUI ’21})}. \bibinfo{publisher}{Association for Computing Machinery}, \bibinfo{address}{New York, NY, USA}, \bibinfo{pages}{1–4}.
\newblock
\showISBNx{978-1-4503-8998-3}
\urldef\tempurl%
\url{https://doi.org/10.1145/3469595.3469607}
\showDOI{\tempurl}


\bibitem[Qiu and Benbasat(2009)]%
        {qiu2009evaluating}
\bibfield{author}{\bibinfo{person}{Lingyun Qiu} {and} \bibinfo{person}{Izak Benbasat}.} \bibinfo{year}{2009}\natexlab{}.
\newblock \showarticletitle{Evaluating anthropomorphic product recommendation agents: A social relationship perspective to designing information systems}.
\newblock \bibinfo{journal}{\emph{Journal of management information systems}} \bibinfo{volume}{25}, \bibinfo{number}{4} (\bibinfo{year}{2009}), \bibinfo{pages}{145--182}.
\newblock


\bibitem[Reicherts et~al\mbox{.}(2022)]%
        {reicherts2022extending}
\bibfield{author}{\bibinfo{person}{Leon Reicherts}, \bibinfo{person}{Gun~Woo Park}, {and} \bibinfo{person}{Yvonne Rogers}.} \bibinfo{year}{2022}\natexlab{}.
\newblock \showarticletitle{Extending Chatbots to probe users: Enhancing complex decision-making through probing conversations}. In \bibinfo{booktitle}{\emph{Proceedings of the 4th Conference on Conversational User Interfaces}}. \bibinfo{pages}{1--10}.
\newblock


\bibitem[Reimer(1996)]%
        {Reimer_1996}
\bibfield{author}{\bibinfo{person}{M. Reimer}.} \bibinfo{year}{1996}\natexlab{}.
\newblock \showarticletitle{The Problem of Dead Metaphors}.
\newblock \bibinfo{journal}{\emph{Philosophical Studies: An International Journal for Philosophy in the Analytic Tradition}} \bibinfo{volume}{82}, \bibinfo{number}{1} (\bibinfo{year}{1996}), \bibinfo{pages}{13–25}.
\newblock
\showISSN{0031-8116}


\bibitem[Sadek et~al\mbox{.}(2023)]%
        {Sadek_Calvo_Mougenot_2023}
\bibfield{author}{\bibinfo{person}{Malak Sadek}, \bibinfo{person}{Rafael~A Calvo}, {and} \bibinfo{person}{Celine Mougenot}.} \bibinfo{year}{2023}\natexlab{}.
\newblock \showarticletitle{Trends, Challenges and Processes in Conversational Agent Design: Exploring Practitioners’ Views through Semi-Structured Interviews}. In \bibinfo{booktitle}{\emph{Proceedings of the 5th International Conference on Conversational User Interfaces}} \emph{(\bibinfo{series}{CUI ’23})}. \bibinfo{publisher}{Association for Computing Machinery}, \bibinfo{address}{New York, NY, USA}, \bibinfo{pages}{1–10}.
\newblock
\showISBNx{9798400700149}
\urldef\tempurl%
\url{https://doi.org/10.1145/3571884.3597143}
\showDOI{\tempurl}


\bibitem[Schiller and McMahon(2019)]%
        {Schiller_McMahon_2019}
\bibfield{author}{\bibinfo{person}{Amy Schiller} {and} \bibinfo{person}{John McMahon}.} \bibinfo{year}{2019}\natexlab{}.
\newblock \showarticletitle{Alexa, Alert Me When the Revolution Comes: Gender, Affect, and Labor in the Age of Home-Based Artificial Intelligence}.
\newblock \bibinfo{journal}{\emph{New Political Science}} \bibinfo{volume}{41}, \bibinfo{number}{2} (\bibinfo{date}{April} \bibinfo{year}{2019}), \bibinfo{pages}{173–191}.
\newblock
\showISSN{0739-3148}
\urldef\tempurl%
\url{https://doi.org/10.1080/07393148.2019.1595288}
\showDOI{\tempurl}


\bibitem[Schuetzler et~al\mbox{.}(2018)]%
        {schuetzler2018influence}
\bibfield{author}{\bibinfo{person}{Ryan~M Schuetzler}, \bibinfo{person}{G~Mark Grimes}, \bibinfo{person}{Justin~Scott Giboney}, {and} \bibinfo{person}{Jay~F Nunamaker~Jr}.} \bibinfo{year}{2018}\natexlab{}.
\newblock \showarticletitle{The influence of conversational agents on socially desirable responding}. In \bibinfo{booktitle}{\emph{Proceedings of the 51st Hawaii International Conference on System Sciences}}. \bibinfo{pages}{283}.
\newblock


\bibitem[Sciuto et~al\mbox{.}(2018)]%
        {Sciuto_Saini_Forlizzi_Hong_2018}
\bibfield{author}{\bibinfo{person}{Alex Sciuto}, \bibinfo{person}{Arnita Saini}, \bibinfo{person}{Jodi Forlizzi}, {and} \bibinfo{person}{Jason~I. Hong}.} \bibinfo{year}{2018}\natexlab{}.
\newblock \showarticletitle{“Hey Alexa, What’s Up?”: A Mixed-Methods Studies of In-Home Conversational Agent Usage}. In \bibinfo{booktitle}{\emph{Proceedings of the 2018 Designing Interactive Systems Conference}} \emph{(\bibinfo{series}{DIS ’18})}. \bibinfo{publisher}{ACM}, \bibinfo{address}{New York, NY, USA}, \bibinfo{pages}{857–868}.
\newblock
\showISBNx{978-1-4503-5198-0}
\urldef\tempurl%
\url{https://doi.org/10.1145/3196709.3196772}
\showDOI{\tempurl}
\newblock
\shownote{00000 event-place: Hong Kong, China}.


\bibitem[Seaborn and Urakami(2021)]%
        {Seaborn_Urakami_2021}
\bibfield{author}{\bibinfo{person}{Katie Seaborn} {and} \bibinfo{person}{Jacqueline Urakami}.} \bibinfo{year}{2021}\natexlab{}.
\newblock \showarticletitle{Measuring Voice UX Quantitatively: A Rapid Review}. In \bibinfo{booktitle}{\emph{Extended Abstracts of the 2021 CHI Conference on Human Factors in Computing Systems}} \emph{(\bibinfo{series}{CHI EA ’21})}. \bibinfo{publisher}{Association for Computing Machinery}, \bibinfo{address}{New York, NY, USA}, \bibinfo{pages}{1–8}.
\newblock
\showISBNx{978-1-4503-8095-9}
\urldef\tempurl%
\url{https://doi.org/10.1145/3411763.3451712}
\showDOI{\tempurl}


\bibitem[Simpson and Crone(2022)]%
        {Simpson_Crone_2022}
\bibfield{author}{\bibinfo{person}{James Simpson} {and} \bibinfo{person}{Cassandra Crone}.} \bibinfo{year}{2022}\natexlab{}.
\newblock \showarticletitle{Should Alexa be a Police Officer, a Doctor, or a Priest?: Towards CUI Relationships Worth Having}. In \bibinfo{booktitle}{\emph{Proceedings of the 4th Conference on Conversational User Interfaces}}. \bibinfo{publisher}{ACM}, \bibinfo{address}{Glasgow United Kingdom}, \bibinfo{pages}{1–5}.
\newblock
\showISBNx{978-1-4503-9739-1}
\urldef\tempurl%
\url{https://doi.org/10.1145/3543829.3544522}
\showDOI{\tempurl}


\bibitem[Stokes et~al\mbox{.}(2004)]%
        {Stokes_Dixon-Woods_McKinley_2004}
\bibfield{author}{\bibinfo{person}{Tim Stokes}, \bibinfo{person}{Mary Dixon-Woods}, {and} \bibinfo{person}{Robert~K McKinley}.} \bibinfo{year}{2004}\natexlab{}.
\newblock \showarticletitle{Ending the doctor–patient relationship in general practice: a proposed model}.
\newblock \bibinfo{journal}{\emph{Family Practice}} \bibinfo{volume}{21}, \bibinfo{number}{5} (\bibinfo{date}{Oct.} \bibinfo{year}{2004}), \bibinfo{pages}{507–514}.
\newblock
\showISSN{0263-2136}
\urldef\tempurl%
\url{https://doi.org/10.1093/fampra/cmh506}
\showDOI{\tempurl}


\bibitem[Tudor~Car et~al\mbox{.}(2020)]%
        {tudor2020conversational}
\bibfield{author}{\bibinfo{person}{Lorainne Tudor~Car}, \bibinfo{person}{Dhakshenya~Ardhithy Dhinagaran}, \bibinfo{person}{Bhone~Myint Kyaw}, \bibinfo{person}{Tobias Kowatsch}, \bibinfo{person}{Shafiq Joty}, \bibinfo{person}{Yin-Leng Theng}, {and} \bibinfo{person}{Rifat Atun}.} \bibinfo{year}{2020}\natexlab{}.
\newblock \showarticletitle{Conversational agents in health care: scoping review and conceptual analysis}.
\newblock \bibinfo{journal}{\emph{Journal of medical Internet research}} \bibinfo{volume}{22}, \bibinfo{number}{8} (\bibinfo{year}{2020}), \bibinfo{pages}{e17158}.
\newblock


\bibitem[Turk(2016)]%
        {Turk_2016}
\bibfield{author}{\bibinfo{person}{Victoria Turk}.} \bibinfo{year}{2016}\natexlab{}.
\newblock \showarticletitle{Home invasion}.
\newblock \bibinfo{journal}{\emph{New Scientist}} \bibinfo{volume}{232}, \bibinfo{number}{3104} (\bibinfo{date}{Dec.} \bibinfo{year}{2016}), \bibinfo{pages}{16–17}.
\newblock
\showISSN{0262-4079}
\urldef\tempurl%
\url{https://doi.org/10.1016/S0262-4079(16)32318-1}
\showDOI{\tempurl}


\bibitem[Vaidyam et~al\mbox{.}(2019)]%
        {vaidyam2019chatbots}
\bibfield{author}{\bibinfo{person}{Aditya~Nrusimha Vaidyam}, \bibinfo{person}{Hannah Wisniewski}, \bibinfo{person}{John~David Halamka}, \bibinfo{person}{Matcheri~S Kashavan}, {and} \bibinfo{person}{John~Blake Torous}.} \bibinfo{year}{2019}\natexlab{}.
\newblock \showarticletitle{Chatbots and conversational agents in mental health: a review of the psychiatric landscape}.
\newblock \bibinfo{journal}{\emph{The Canadian Journal of Psychiatry}} \bibinfo{volume}{64}, \bibinfo{number}{7} (\bibinfo{year}{2019}), \bibinfo{pages}{456--464}.
\newblock


\bibitem[van Heerden et~al\mbox{.}(2017)]%
        {van2017potential}
\bibfield{author}{\bibinfo{person}{Alastair van Heerden}, \bibinfo{person}{Xolani Ntinga}, {and} \bibinfo{person}{Khanya Vilakazi}.} \bibinfo{year}{2017}\natexlab{}.
\newblock \showarticletitle{The potential of conversational agents to provide a rapid HIV counseling and testing services}. In \bibinfo{booktitle}{\emph{2017 international conference on the frontiers and advances in data science (FADS)}}. IEEE, \bibinfo{pages}{80--85}.
\newblock


\bibitem[V{\"o}lkel et~al\mbox{.}(2021)]%
        {volkel2021manipulating}
\bibfield{author}{\bibinfo{person}{Sarah~Theres V{\"o}lkel}, \bibinfo{person}{Samantha Meindl}, {and} \bibinfo{person}{Heinrich Hussmann}.} \bibinfo{year}{2021}\natexlab{}.
\newblock \showarticletitle{Manipulating and evaluating levels of personality perceptions of voice assistants through enactment-based dialogue design}. In \bibinfo{booktitle}{\emph{Proceedings of the 3rd Conference on Conversational User Interfaces}}. \bibinfo{pages}{1--12}.
\newblock


\bibitem[Wang et~al\mbox{.}(2020)]%
        {Wang_Yang_Shao_Abdullah_Sundar_2020}
\bibfield{author}{\bibinfo{person}{Jinping Wang}, \bibinfo{person}{Hyun Yang}, \bibinfo{person}{Ruosi Shao}, \bibinfo{person}{Saeed Abdullah}, {and} \bibinfo{person}{S.~Shyam Sundar}.} \bibinfo{year}{2020}\natexlab{}.
\newblock \showarticletitle{Alexa as Coach: Leveraging Smart Speakers to Build Social Agents that Reduce Public Speaking Anxiety}. In \bibinfo{booktitle}{\emph{Proceedings of the 2020 CHI Conference on Human Factors in Computing Systems}} \emph{(\bibinfo{series}{CHI ’20})}. \bibinfo{publisher}{Association for Computing Machinery}, \bibinfo{address}{New York, NY, USA}, \bibinfo{pages}{1–13}.
\newblock
\showISBNx{978-1-4503-6708-0}
\urldef\tempurl%
\url{https://doi.org/10.1145/3313831.3376561}
\showDOI{\tempurl}


\bibitem[Wei et~al\mbox{.}(2023)]%
        {Wei_Kim_Kuzminykh_2023}
\bibfield{author}{\bibinfo{person}{Christina~Ziying Wei}, \bibinfo{person}{Young-Ho Kim}, {and} \bibinfo{person}{Anastasia Kuzminykh}.} \bibinfo{year}{2023}\natexlab{}.
\newblock \showarticletitle{The Bot on Speaking Terms: The Effects of Conversation Architecture on Perceptions of Conversational Agents}. In \bibinfo{booktitle}{\emph{Proceedings of the 5th International Conference on Conversational User Interfaces}} \emph{(\bibinfo{series}{CUI ’23})}. \bibinfo{publisher}{Association for Computing Machinery}, \bibinfo{address}{New York, NY, USA}, \bibinfo{pages}{1–16}.
\newblock
\showISBNx{9798400700149}
\urldef\tempurl%
\url{https://doi.org/10.1145/3571884.3597139}
\showDOI{\tempurl}


\bibitem[Yang et~al\mbox{.}(2019)]%
        {Yang_Aurisicchio_Baxter_2019}
\bibfield{author}{\bibinfo{person}{Xi Yang}, \bibinfo{person}{Marco Aurisicchio}, {and} \bibinfo{person}{Weston Baxter}.} \bibinfo{year}{2019}\natexlab{}.
\newblock \showarticletitle{Understanding Affective Experiences with Conversational Agents}. In \bibinfo{booktitle}{\emph{Proceedings of the 2019 CHI Conference on Human Factors in Computing Systems}} \emph{(\bibinfo{series}{CHI ’19})}. \bibinfo{publisher}{ACM}, \bibinfo{address}{New York, NY, USA}, \bibinfo{pages}{542:1--542:12}.
\newblock
\showISBNx{978-1-4503-5970-2}
\urldef\tempurl%
\url{https://doi.org/10.1145/3290605.3300772}
\showDOI{\tempurl}
\newblock
\shownote{00000 event-place: Glasgow, Scotland Uk}.


\bibitem[Zhu et~al\mbox{.}(2023)]%
        {Zhu_Pysander_Soderberg_2023}
\bibfield{author}{\bibinfo{person}{Hui Zhu}, \bibinfo{person}{Eva-Lotta Pysander}, {and} \bibinfo{person}{Ingalill Soderberg}.} \bibinfo{year}{2023}\natexlab{}.
\newblock \showarticletitle{Not transparent and incomprehensible: A qualitative user study of an AI-empowered financial advisory system}.
\newblock \bibinfo{journal}{\emph{Data and Information Management}}  \bibinfo{volume}{7} (\bibinfo{date}{April} \bibinfo{year}{2023}), \bibinfo{pages}{100041}.
\newblock
\urldef\tempurl%
\url{https://doi.org/10.1016/j.dim.2023.100041}
\showDOI{\tempurl}


\end{thebibliography}

\onecolumn

\end{document}